\documentclass[conference]{IEEEtran}
\IEEEoverridecommandlockouts
\usepackage{cite}
\usepackage{amsmath,amssymb,amsfonts}
\usepackage{algorithmic}
\usepackage{graphicx}
\usepackage{textcomp}
\usepackage{xcolor}
\usepackage{hyperref}

\usepackage{amsmath,graphicx,tikz,multirow,xcolor,balance, arydshln}

\usetikzlibrary{decorations.pathreplacing, positioning, calc}

\def\BibTeX{{\rm B\kern-.05em{\sc i\kern-.025em b}\kern-.08em
    T\kern-.1667em\lower.7ex\hbox{E}\kern-.125emX}}

\makeatletter
\def\ps@IEEEtitlepagestyle{%
	\def\@oddfoot{\mycopyrightnotice}%
	\def\@oddhead{\hbox{}\@IEEEheaderstyle\leftmark\hfil\thepage}\relax
	\def\@evenhead{\@IEEEheaderstyle\thepage\hfil\leftmark\hbox{}}\relax
	\def\@evenfoot{}%
}
\def\mycopyrightnotice{%
	\begin{minipage}{\textwidth}
		\scriptsize
		\copyright 2024 IEEE. Personal use of this material is permitted. Permission from IEEE must be obtained for all other uses, in any current or future media, including reprinting/republishing this material for advertising or promotional purposes, creating new collective works, for resale or redistribution to servers or lists, or reuse of any copyrighted component of this work in other works. DOI: \url{10.1109/MMSP61759.2024.10743810} (MMSP 2024)
	\end{minipage}
}
\makeatother

\begin{document}

\title{Inter-Camera Color Correction for Multispectral Imaging with Camera Arrays Using a Consensus Image
\thanks{The authors gratefully acknowledge that this work has been supported by the Bayrische Forschungsstiftung (BFS, Bavarian Research Foundation) under project number AZ-1547-22.}
}

\author{\IEEEauthorblockN{Katja Kossira, J\"urgen Seiler, and Andr\'e Kaup}
\IEEEauthorblockA{\textit{Multimedia Communications and Signal Processing} \\
\textit{Friedrich-Alexander University Erlangen-N\"urnberg (FAU)}\\
Erlangen, Germany \\
\{katja.kossira, juergen.seiler, andre.kaup \} @fau.de}
}

\maketitle

\begin{abstract}
This paper introduces a novel method for inter-camera color calibration for multispectral imaging with camera arrays using a consensus image. Capturing images using multispectral camera arrays has gained importance in medical, agricultural, and environmental processes. Due to fabrication differences, noise, or device altering, varying pixel sensitivities occur, influencing classification processes. Therefore, color calibration between the cameras is necessary. In existing methods, one of the camera images is chosen and considered as a reference, ignoring the color information of all other recordings. Our new approach does not just take one image as reference, but uses statistical information such as the location parameter to generate a consensus image as basis for calibration. This way, we managed to improve the PSNR values for the linear regression color correction algorithm by 1.15 dB and the improved color difference (iCID) values by 2.81.
\end{abstract}


\section{Introduction}
\label{sec:intro}
Capturing images is an everyday action for almost everyone in today's life. Also the separation of light into its single spectral bands using multiple cameras for multispectral imaging (MSI) setups has proven its usefulness in many classification processes, thus providing possible applications in numerous fields. MSI is based on the fact that diverse substances absorb and reflect light of various spectral ranges differently for specific wavelengths. 

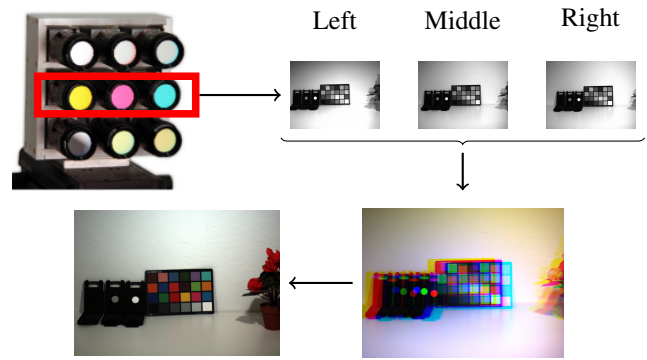
\begin{figure}
\begin{tikzpicture}
	\node (CAMSI) [yshift = -2.1cm, xshift = -1cm] {\includegraphics[scale=0.43]{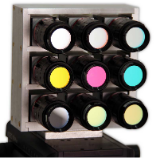}};
	\node(kasten) [draw=red, minimum width = 2.1cm, minimum height = 0.5cm, below of =CAMSI, yshift=1.1cm, line width =0.1cm, xshift =0.2cm]{};

	\node (left)[right of=kasten, xshift=1.9cm]{\includegraphics[scale=0.028]{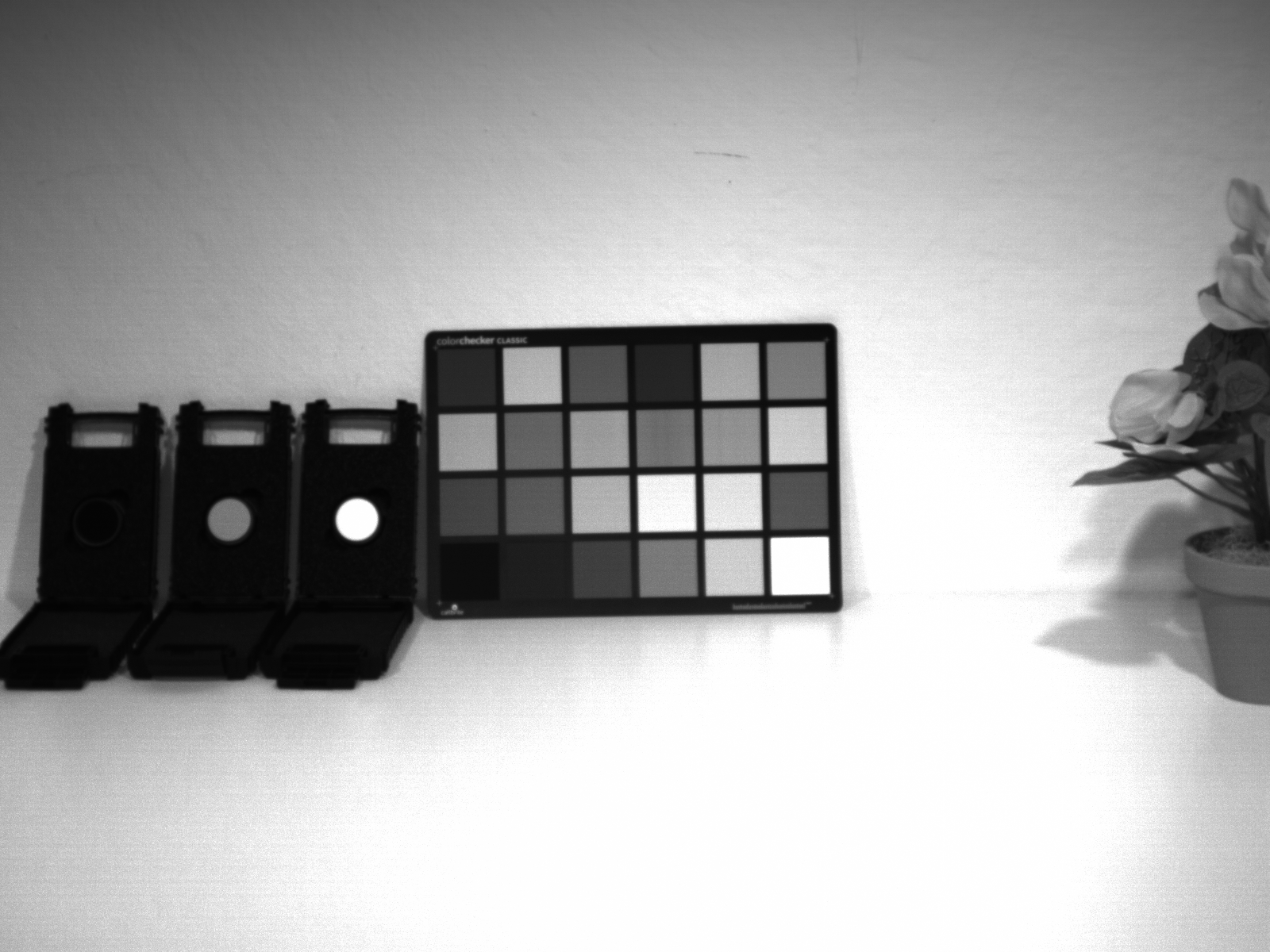}};
	\node(center)[right of=left, xshift=0.7cm]{\includegraphics[scale=0.028]{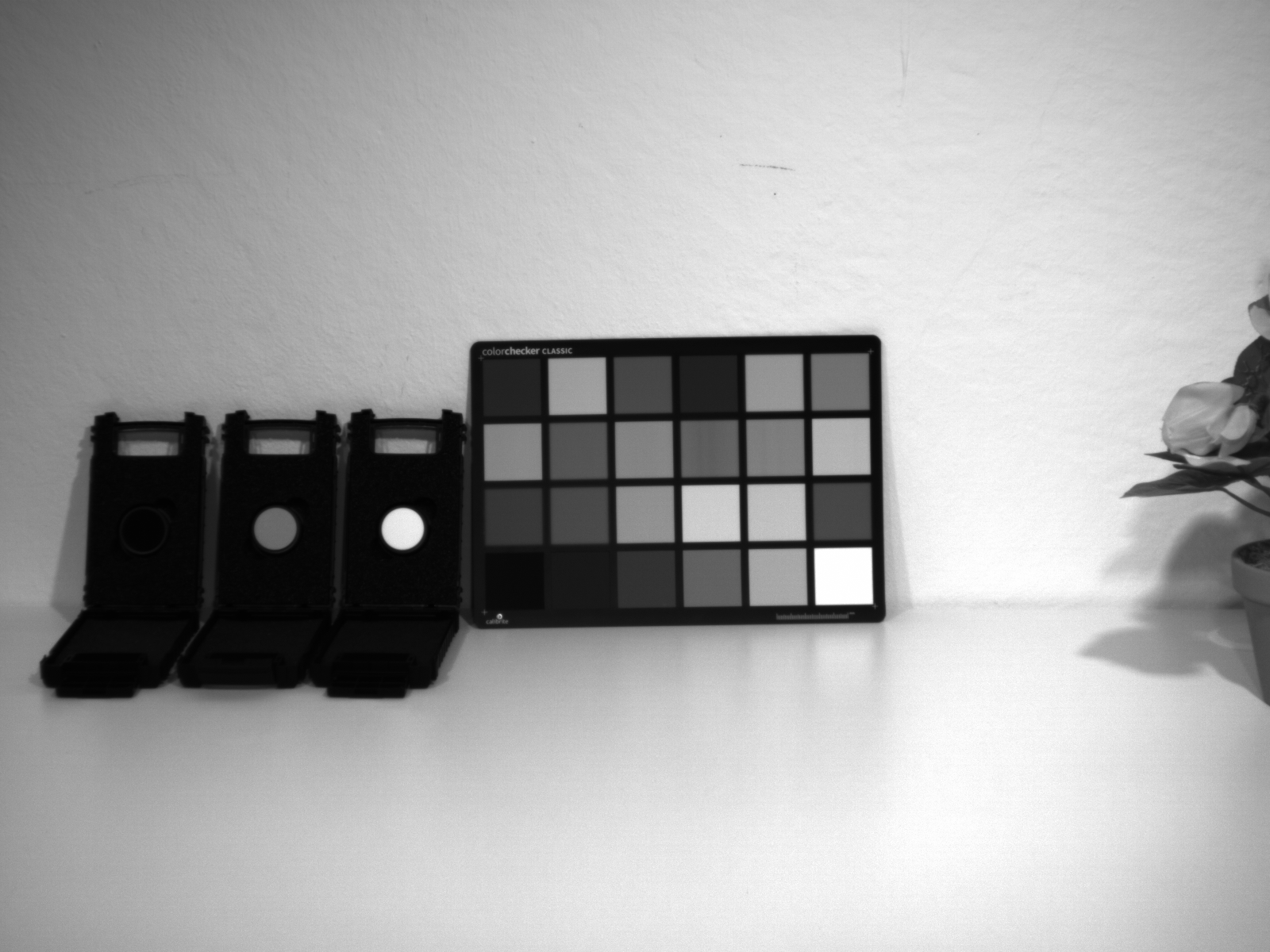}};
	\node(right)[right of=center, xshift=0.7cm]{\includegraphics[scale=0.028]{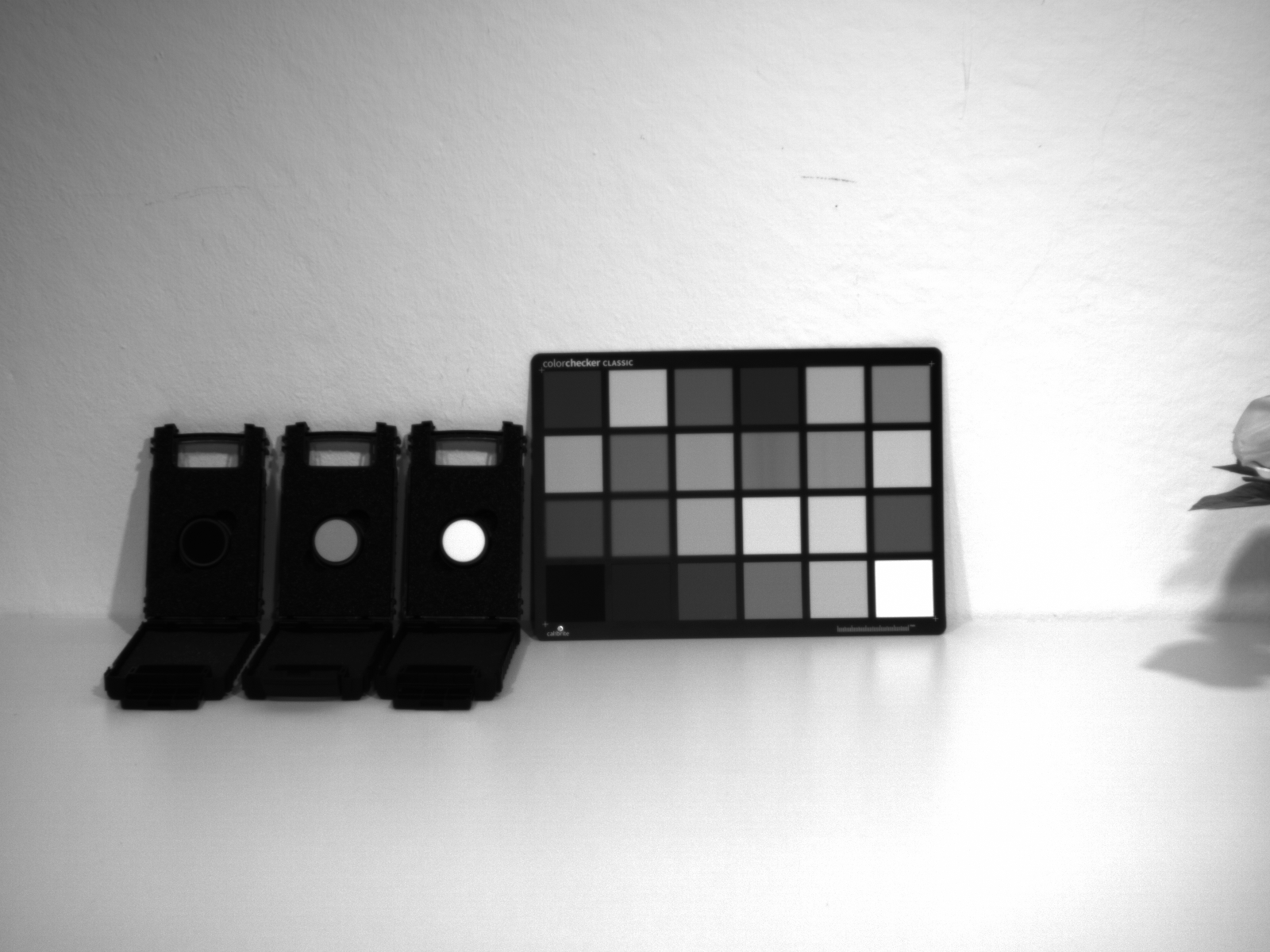}};
	\node(leftdes) [above of = left] {Left};
	\node (centerdes) [above of=center]{Middle};
	\node(rightdes)[above of = right]{Right};
	\draw[->,thick](kasten)--(left);
	\draw [decorate, decoration={brace, mirror}]  (left.south west) -- (right.south east);
	\draw[->,thick] ($(left.south)!0.5!(right.south) + (0,-0.2)$) -- ++(0,-0.5);
	
	\node (overlaid)[below of = center, yshift=-1.5cm]{\includegraphics[scale=0.21]{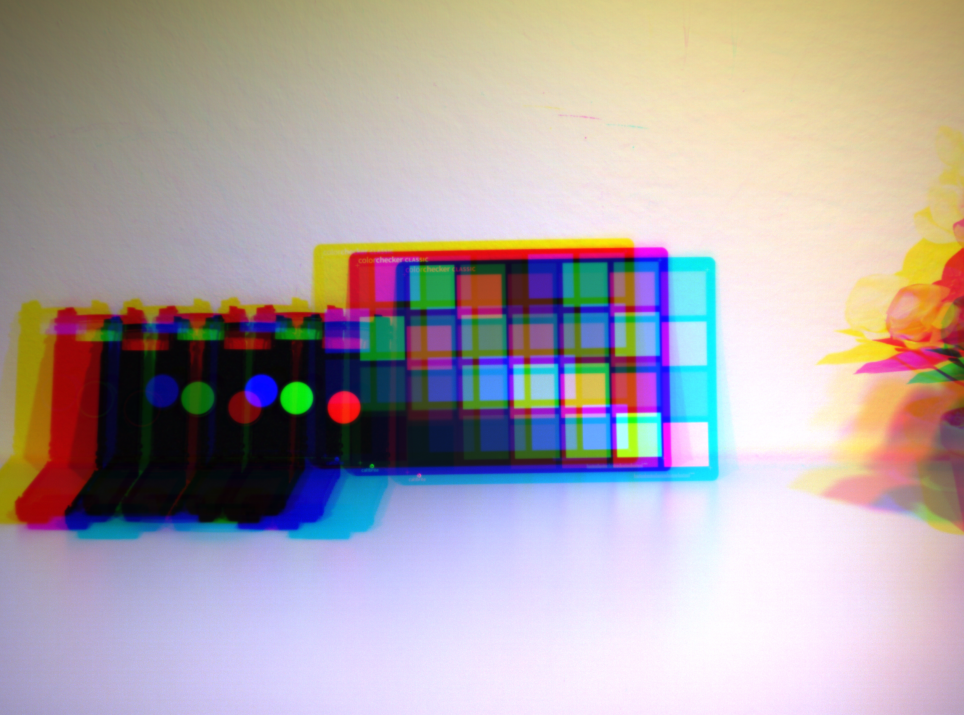}};
	\node (regisrecon)[left of = overlaid, xshift=-2.8cm]{\includegraphics[scale=0.41]{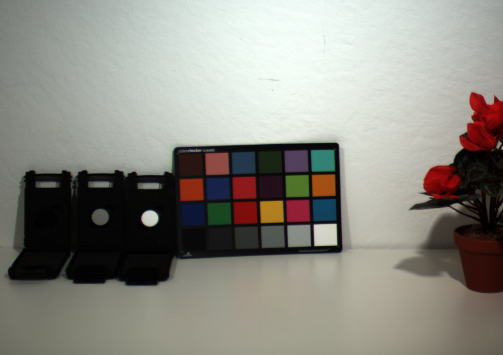}};
	\draw[->,thick](overlaid)--(regisrecon);
	
\end{tikzpicture}
\vspace{-0.3cm}
\caption{A multispectral camera array (CAMSI \cite{CAMSI}) with three corresponding recorded images of the center row (top). Additionally, the overlaid image with horizontal displacement (bottom right) and the registrated and reconstructed image (bottom left)
are depicted.}
\label{fig:CAMSI}
\vspace{-0.8cm}
\end{figure}

One possible setup is the camera array multispectral imaging \cite{CAMSI}, which is shown in Fig. \ref{fig:CAMSI} on the upper left. Different filters of varying wavelengths optimized for the specific tasks \cite{Kossira} can be placed in front of the cameras, allowing for simple individual exchanges depending on the field of application, which include for example measuring the high dynamic range spectral reflectance of artworks \cite{HSIArt},  measuring the total outgoing radiation of the earth for climate protection decisions \cite{MSIRadiation}, or precision agriculture for better crop monitoring and increased food production \cite{MSIAgriculture}.

The cameras are placed at different positions, what results in varying viewing angles of the scene. Fig. \ref{fig:CAMSI}additionally shows three corresponding recorded images of the center row, which are overlaid. By applying rectification, registration and reconstruction algorithms \cite{CAMSI} to the images, a good alignment towards the center camera can be guaranteed as shown in Fig. \ref{fig:CAMSI} in the bottom row and the local displacement is negligible, since missing pixel information is regained.

\begin{figure}[t!]
	\begin{tikzpicture}[      
		every node/.style={anchor=south west,inner sep=20pt},
		x=1mm, y=1mm,
		]   
		\node (fig1) at (0,0)
		{\resizebox{0.48\textwidth}{!}{\input{Images_Paper/histogram_CAMSI_introduction.pgf}}};
		\node (fig2) at (12,28)
		{\includegraphics[scale=0.02]{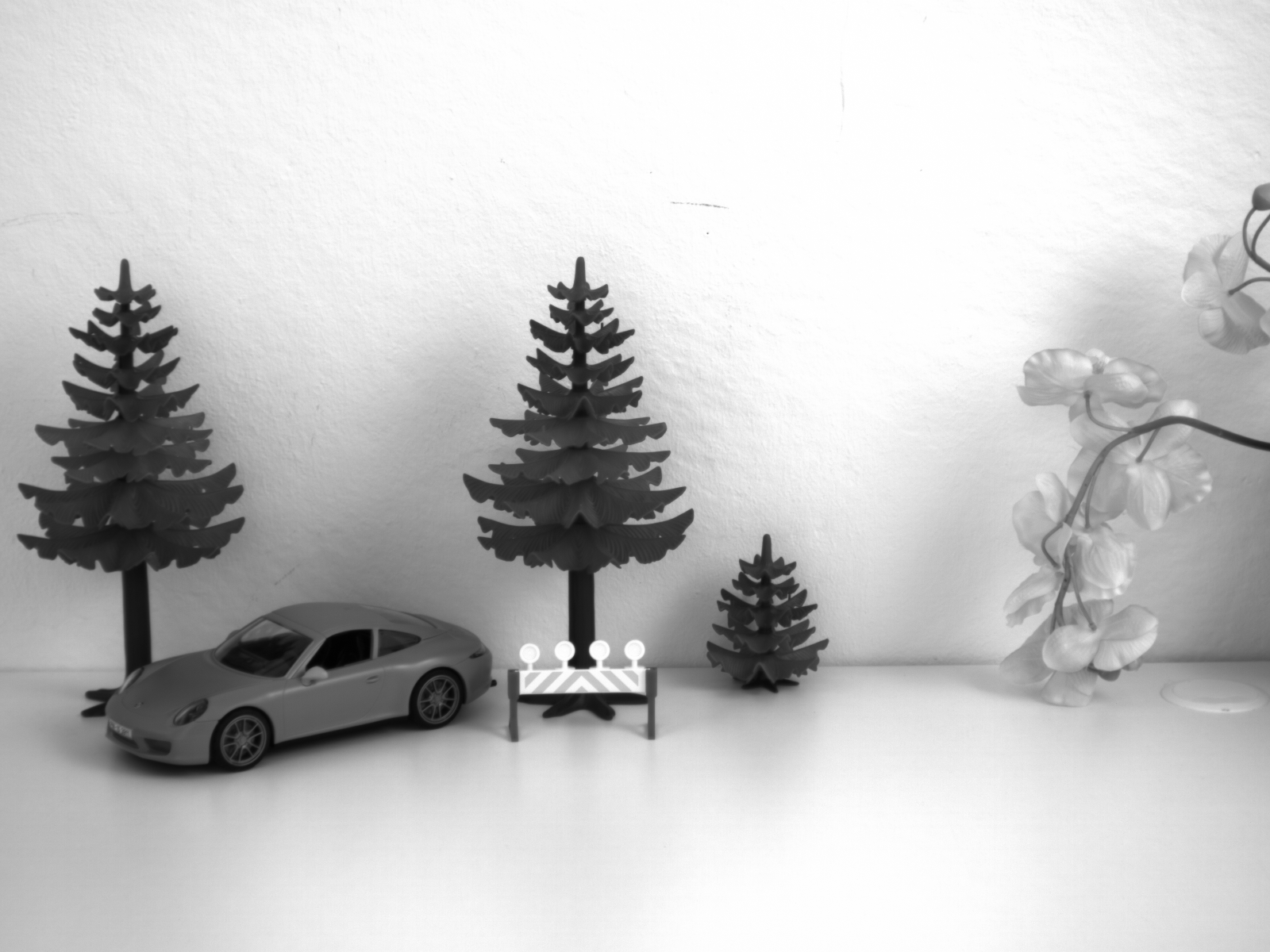}};  
		\node (fig3) [right of=fig2, align=center, xshift=0.2cm]
		{\includegraphics[scale=0.02]{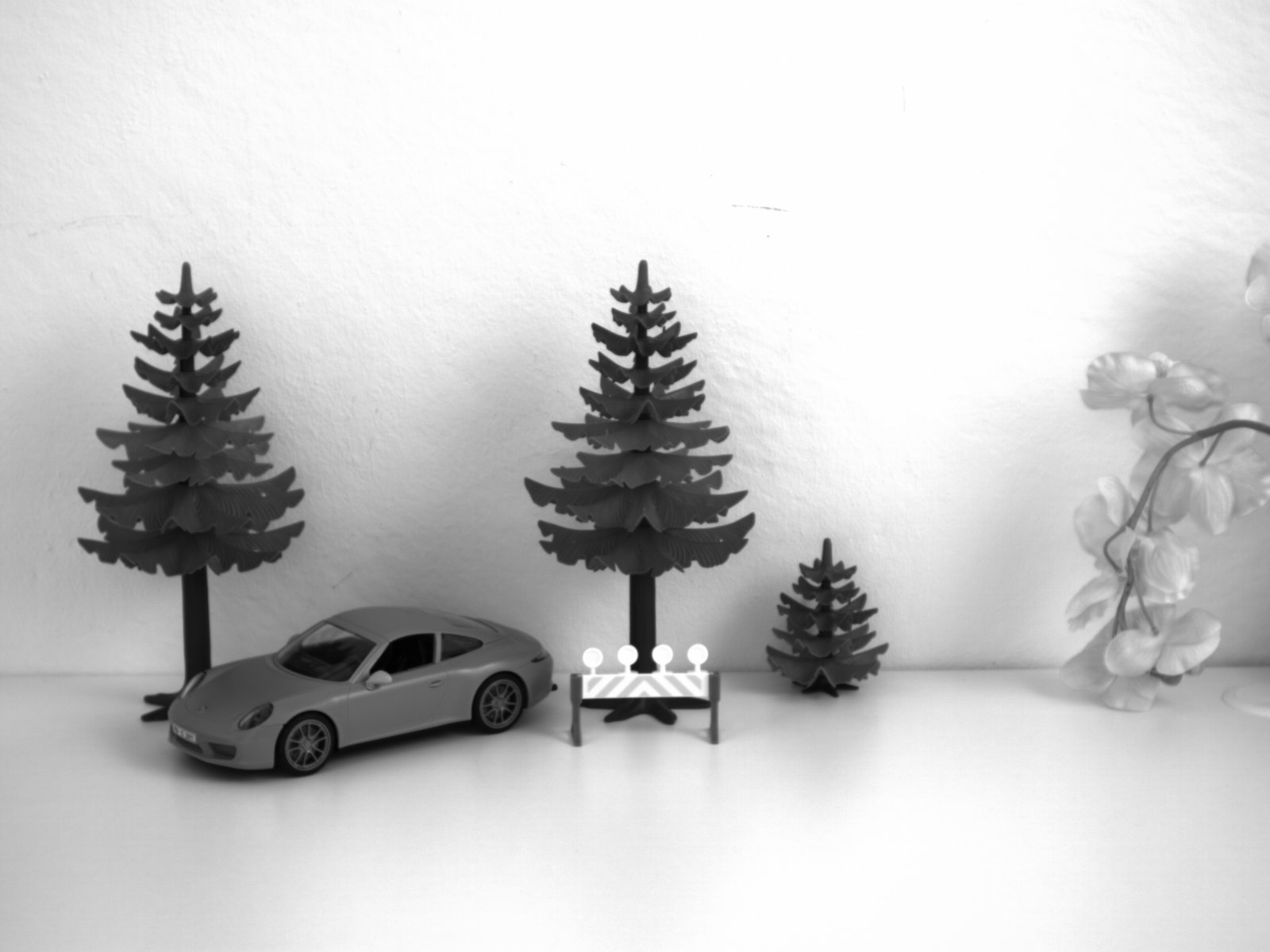}}; 
		\node (fig4) [right of=fig3, align=center, xshift=0.2cm]
		{\includegraphics[scale=0.02]{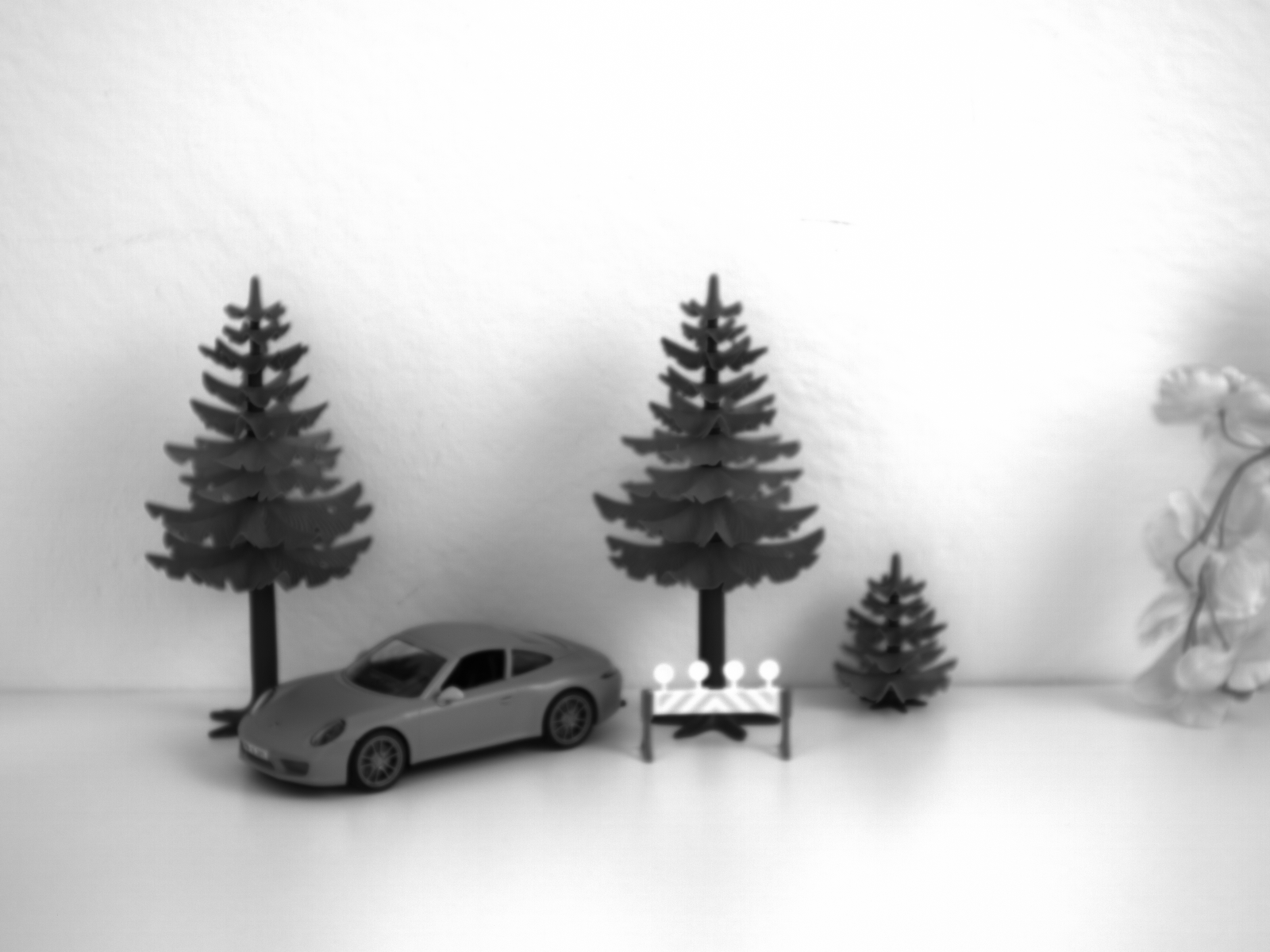}}; 
		\node (fig5) [below of=fig2, align=center, yshift=0.07cm]
		{\includegraphics[scale=0.02]{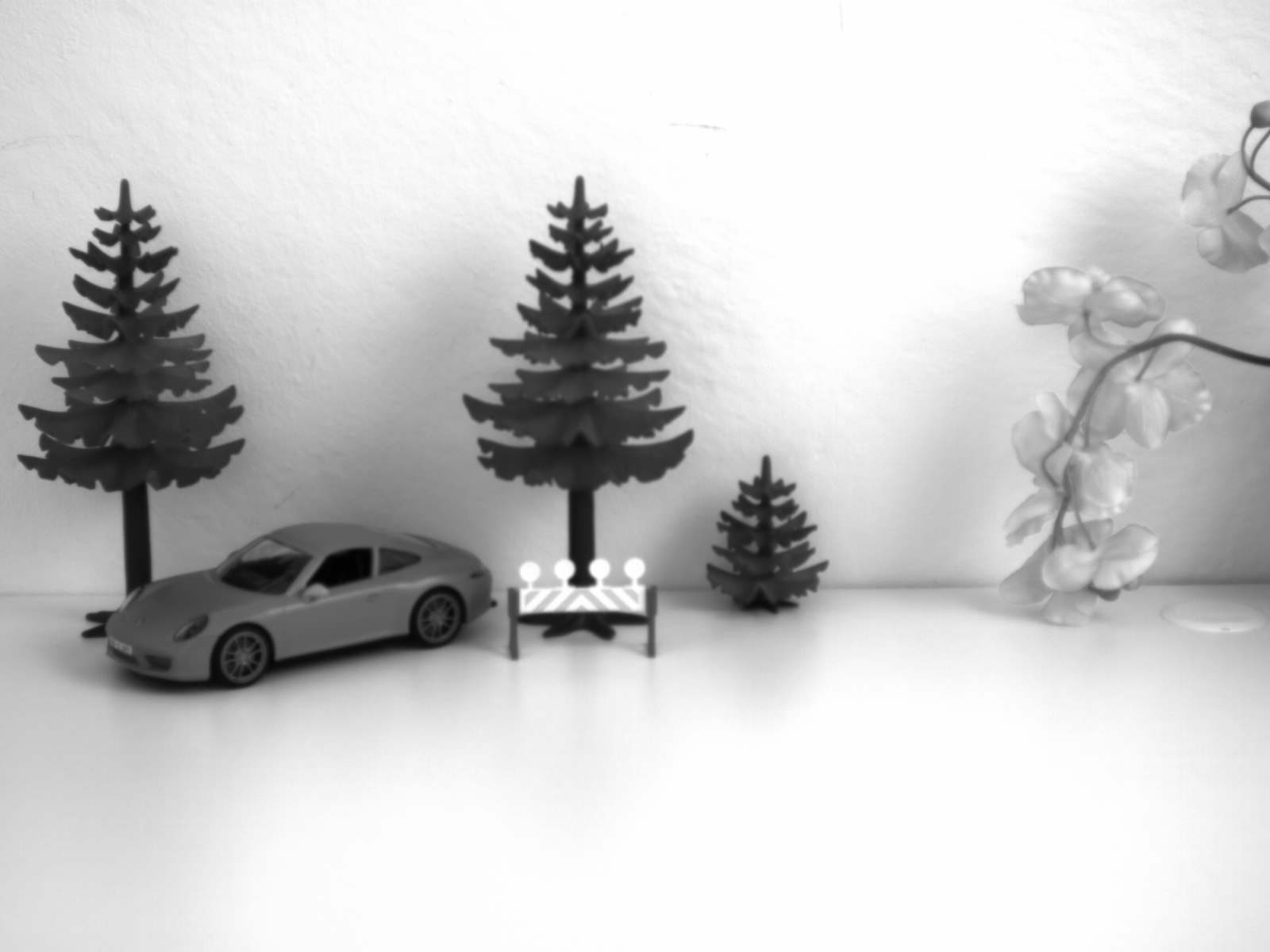}}; 
		\node (fig6) [right of=fig5, align=center, xshift=0.2cm]
		{\includegraphics[scale=0.02]{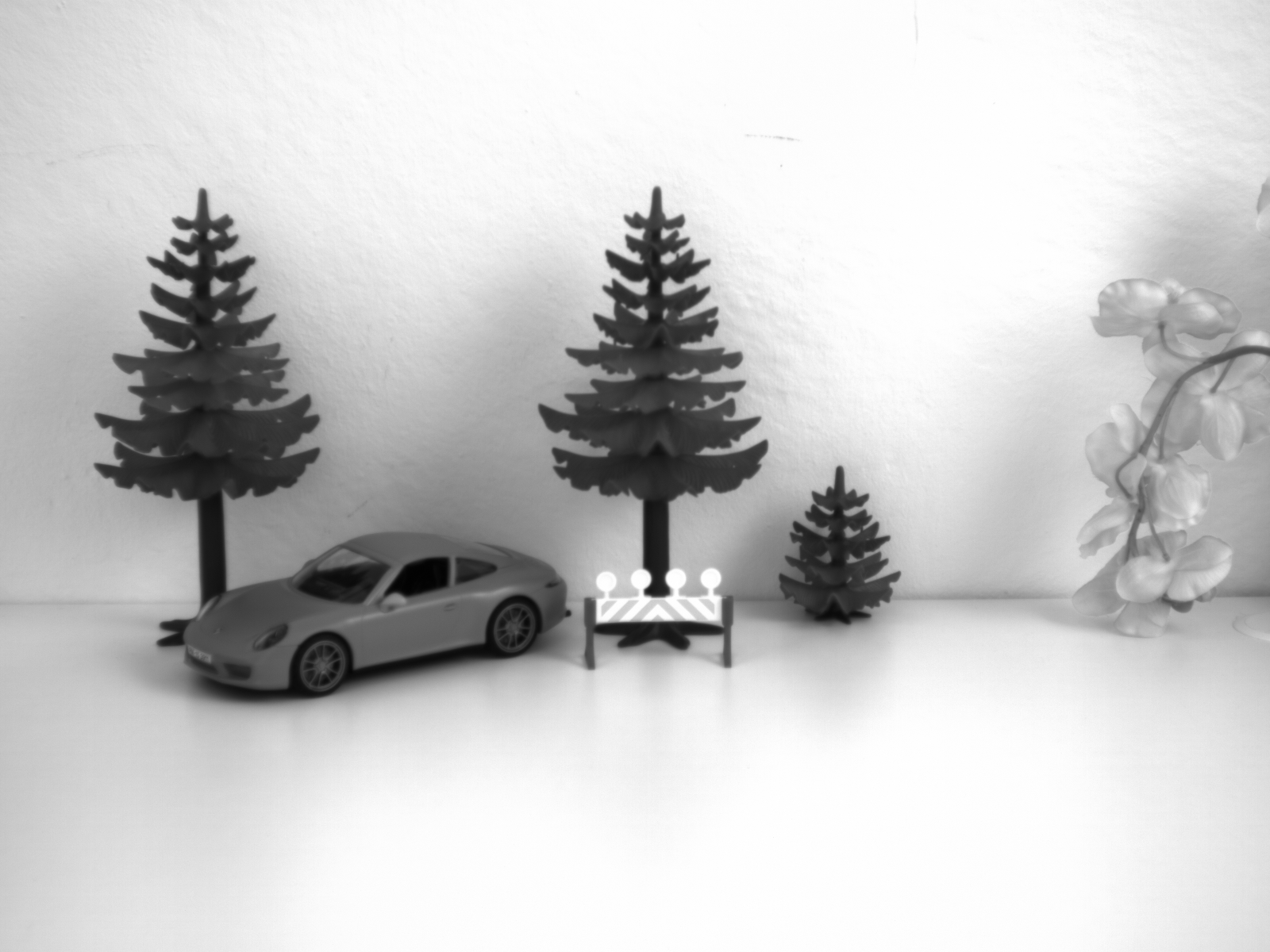}}; \\
		\node (fig7) [right of=fig6, align=center, xshift=0.2cm]
		{\includegraphics[scale=0.02]{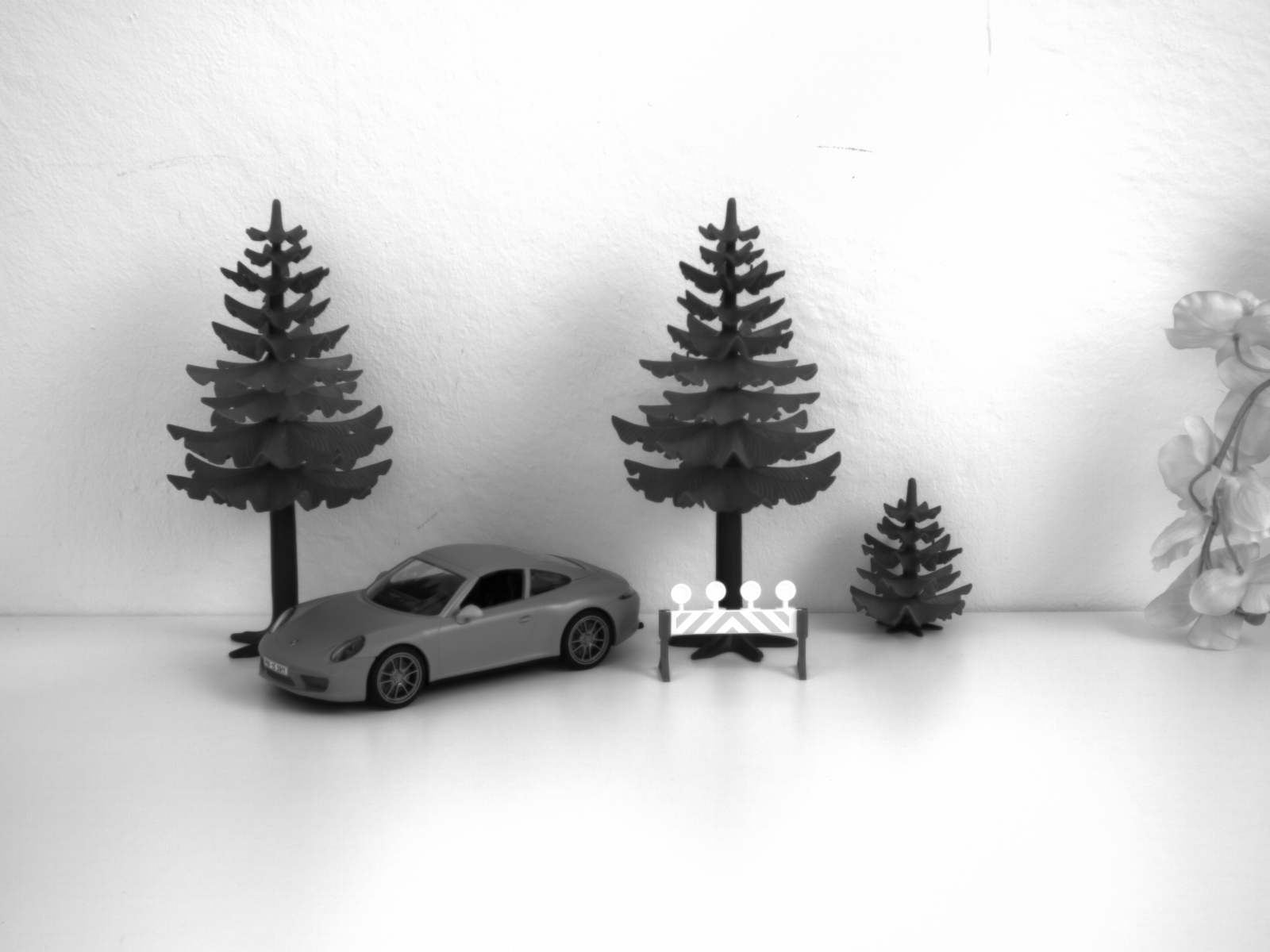}}; 
		\node (fig8) [below of=fig5, align=center, yshift=0.07cm]
		{\includegraphics[scale=0.02]{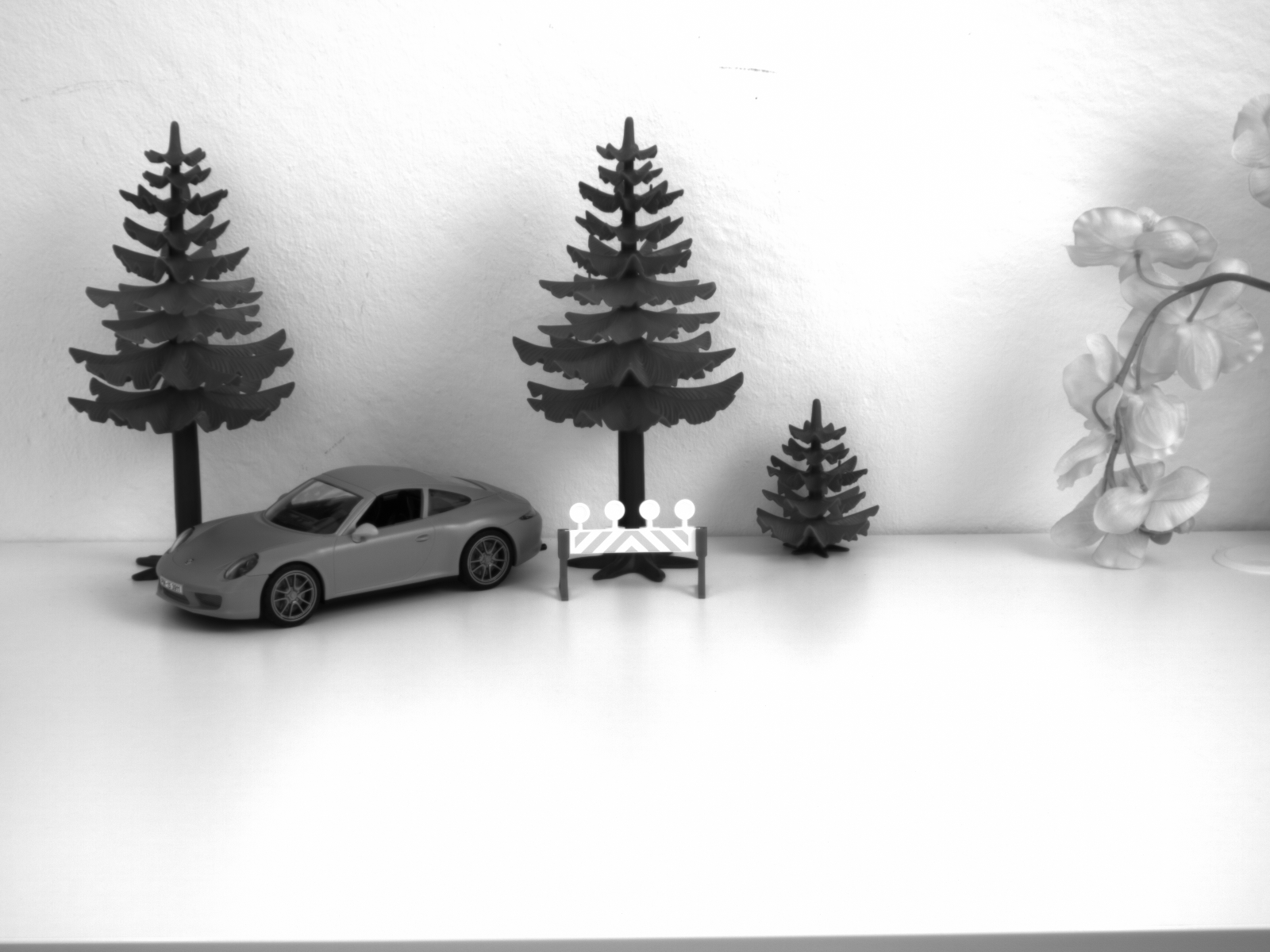}}; \\
		\node (fig9) [right of=fig8, align=center, xshift=0.2cm]
		{\includegraphics[scale=0.02]{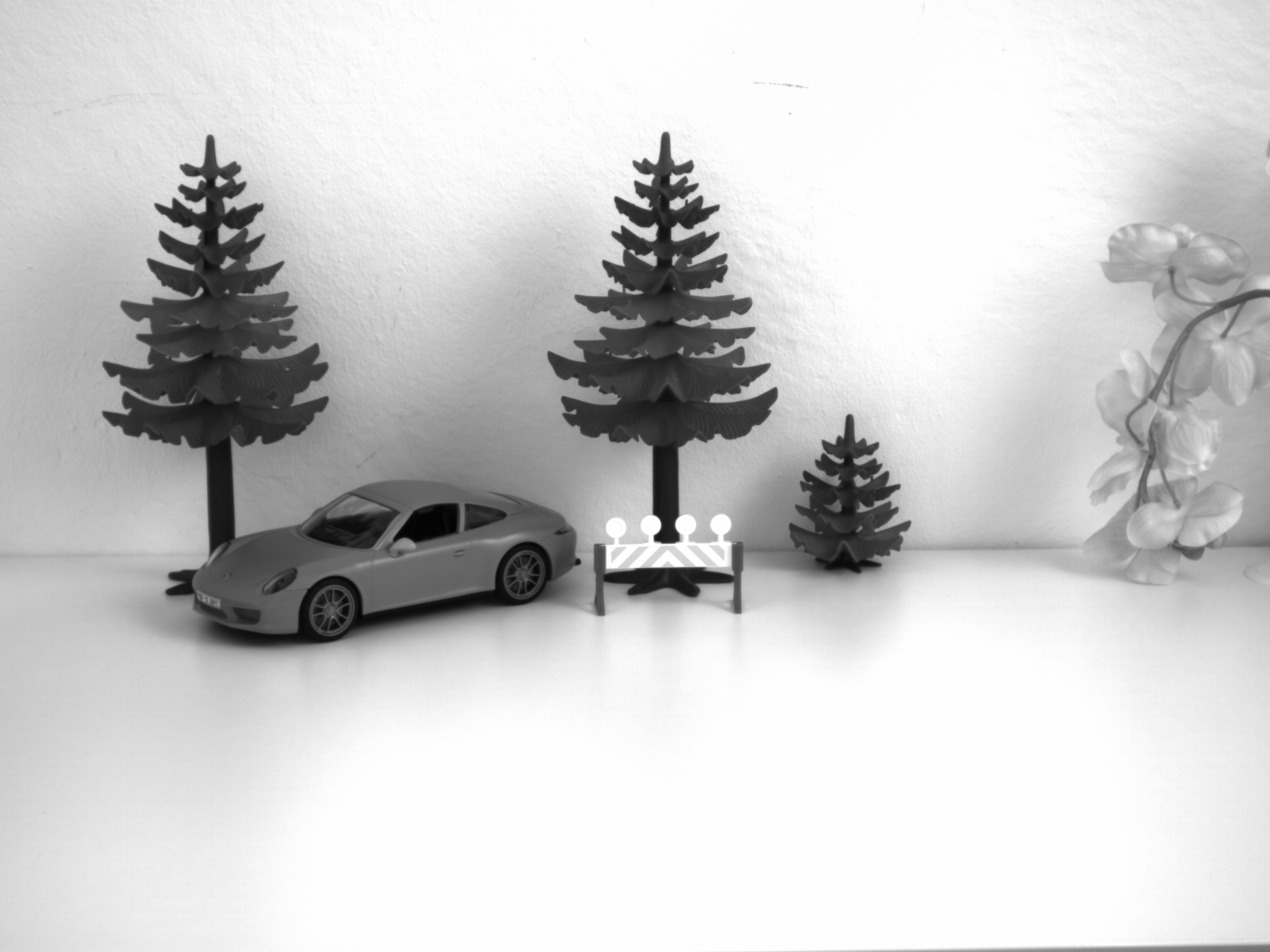}}; \\
		\node (fig10) [right of=fig9, align=center, xshift=0.2cm]
		{\includegraphics[scale=0.02]{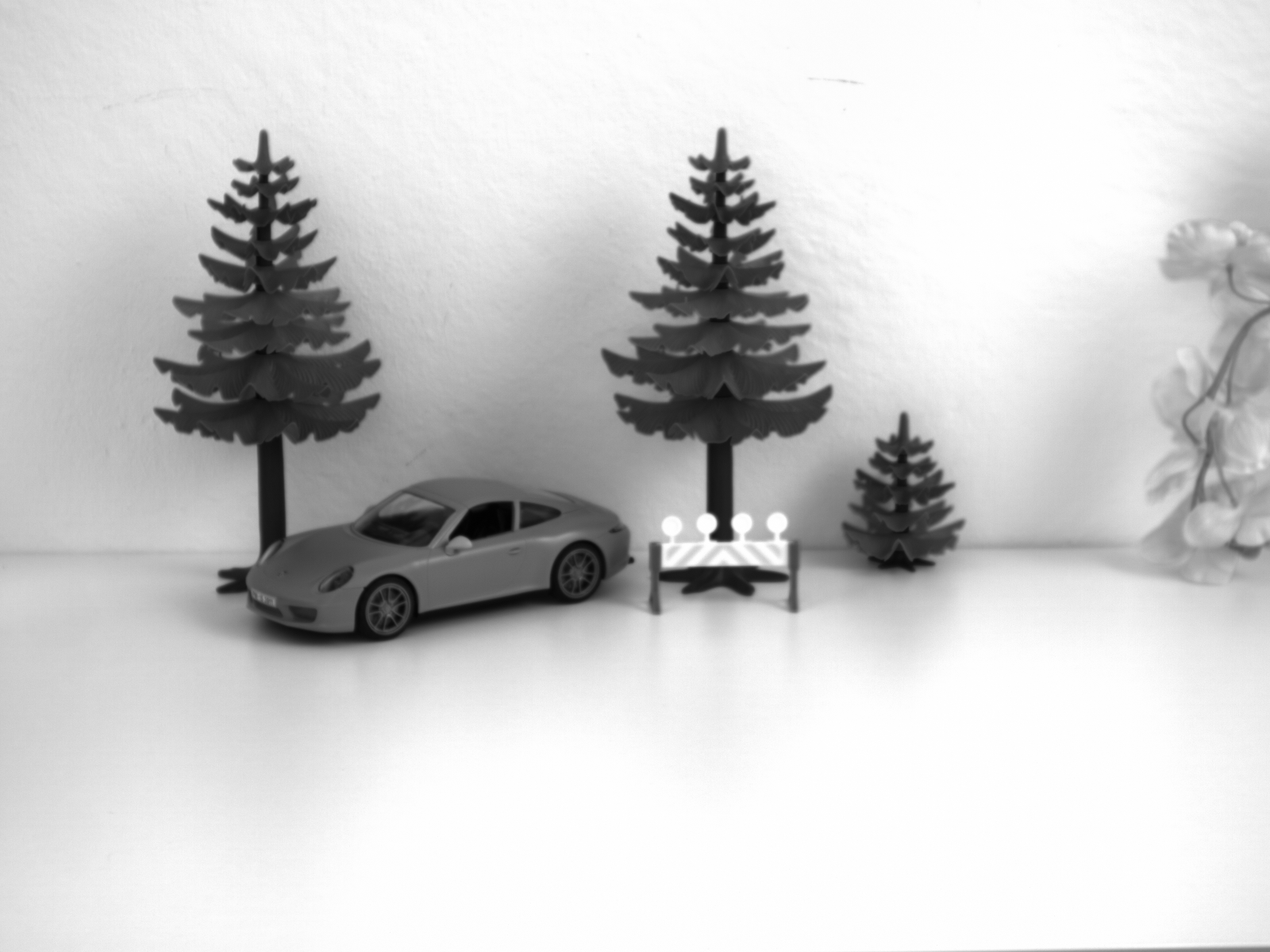}}; \\
	\end{tikzpicture}
	\vspace{-0.8cm}
	\caption{Histograms and recorded scene of the 9 cameras of the multispectral camera array CAMSI \cite{CAMSI} after reconstruction and registration.}
	\label{image color responses CAMSI}
	\vspace{-0.6cm}
\end{figure}

Since MSI applications are primarily used for unambiguous evaluation of varying scenes, consistent color results across the employed cameras are indispensable. However, fabrication differences, device altering, thermal issues or electrical noise can lead to inconsistencies, even though the cameras are of the same type and producer \cite{ensuringColorConsistencyIlie}. These dynamic deviations require calibration algorithms for obtaining consistent results, since the simple solution of calibrating with a gray wedge only ensure measurement accuracy with reality but no inter-camera consistency. Fig. \ref{image color responses CAMSI} illustrates the histograms of the CAMSI setup of the 9 cameras \textit{without} mounted filters of a real scene showing different sorts of plastic recorded simultaneously at the same time. The before mentioned rectification, registration and reconstruction processes have already been performed on these images, hence the different viewing angles between the cameras can be neglected. Even though the images were taken under the same hardware and illumination conditions, the illustration reveals deviating histograms, thus confirming the problem of different pixel sensitivities within the camera setup.

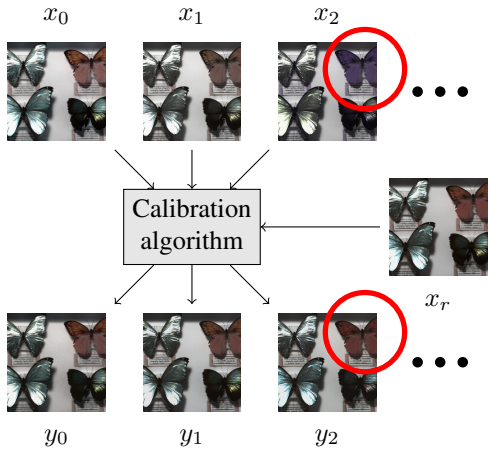
\begin{figure}[t!]
	\begin{center}
		\begin{tikzpicture}[auto]
			\centering
			% Color correction algorithm block
			\node (algorithm) [draw, fill=gray!20, minimum width=1.5cm, minimum height=1cm, align = center] {Calibration\\algorithm};
			
			\node (ref) [right of=algorithm, anchor=west, xshift=1.5cm] {\includegraphics[width=1.3cm, angle=-90]{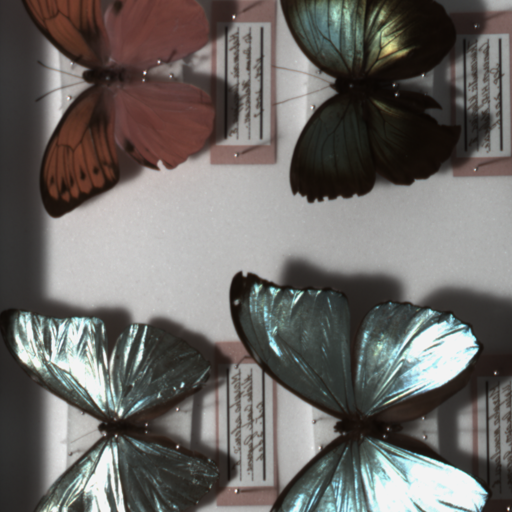}} ;
			\node(labelref) [below of=ref] {$x_{r}$};
			% Input blocks
			%\node (input2) [draw, fill=yellow!20, above of=algorithm, minimum width=0.5cm, minimum height=0.5cm, align = center] {Img2};
			\node (input2) [above of=algorithm, align=center, yshift=0.8cm] {\includegraphics[width=1.3cm, angle=-90]{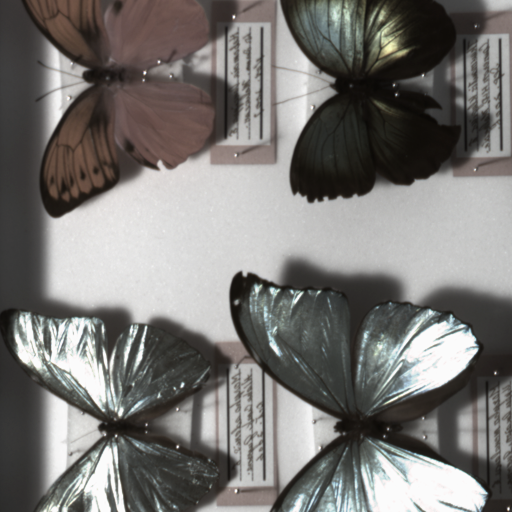}};
			\node (labelin2) [above of=input2] {$x_{1}$};
			%\node (input1) [draw, fill=red!20, left of=input2, minimum width=0.5cm, minimum height=0.5cm, align = left] {Img1};
			\node (input1) [left of=input2, align=center, xshift=-0.8cm] {\includegraphics[width=1.3cm, angle=-90]{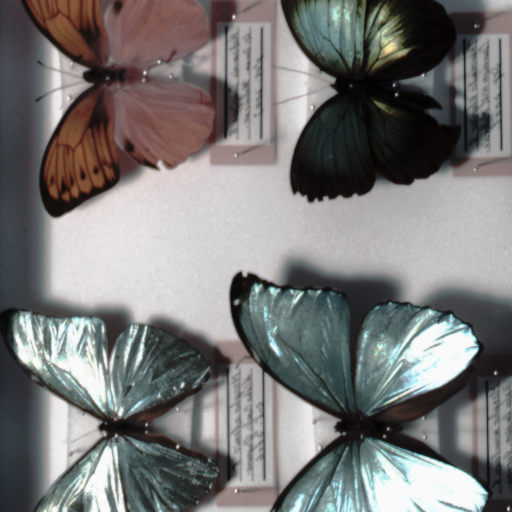}};
			\node (labelin1) [above of=input1] {$x_{0}$};
			%\node (input3) [draw, fill=orange!20,  right of=input2, minimum width=0.5cm, minimum height=0.5cm, align = right] {Img3};
			\node (input3) [right of=input2, align=center, xshift=0.8cm] {\includegraphics[width=1.3cm, angle=-90]{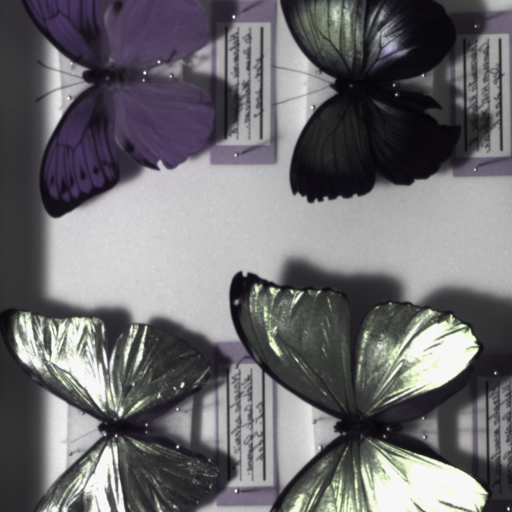}};
			\node (labelin3) [above of=input3] {$x_{2}$};
			% Output blocks
			%\node (output2) [draw, fill=purple!20, below of=algorithm, minimum width=0.5cm, minimum height=0.5cm] {Img2};
			\node (output2) [below of=algorithm, yshift=-0.8cm] {\includegraphics[width=1.3cm, angle=-90]{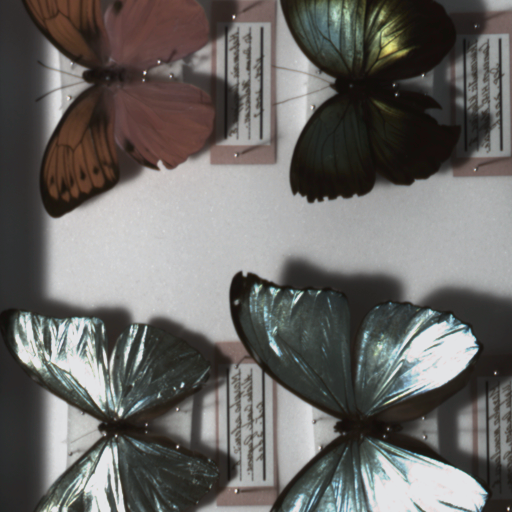}};
			\node(labelout2) [below of=output2] {$y_{1}$};
		%	\node (output1) [draw, fill=purple!20, left of=output2, minimum width=0.5cm, minimum height=0.5cm] {Img1};
			\node (output1) [left of=output2, xshift=-0.8cm] {\includegraphics[width=1.3cm,angle=-90]{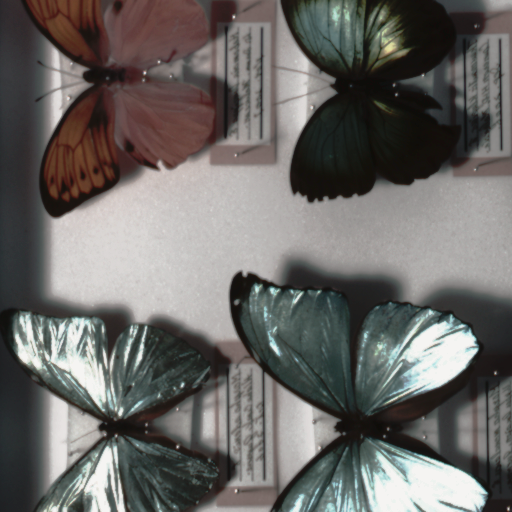}};
			\node (labelout1) [below of=output1] {$y_{0}$};
		%	\node (output3) [draw, fill=purple!20, right of=output2, minimum width=0.5cm, minimum height=0.5cm] {Img3};
			\node (output3) [right of=output2, xshift=0.8cm] {\includegraphics[width=1.3cm, angle=-90]{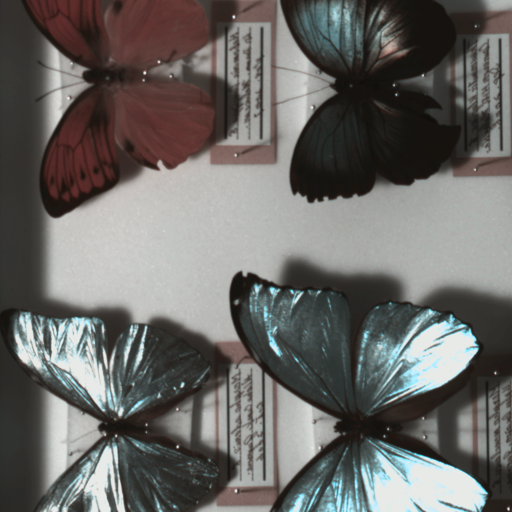}};
			\node (labelout3) [below of=output3] {$y_{2}$};
			% Arrows
			\draw[->] (ref) -- (algorithm);
			\draw[->] (input1) -- (algorithm);
			\draw[->] (input2) -- (algorithm);
			\draw[->] (input3) -- (algorithm);
			\draw[->] (algorithm) -- (output1);
			\draw[->] (algorithm) -- (output2);
			\draw[->] (algorithm) -- (output3);
			
			\draw[red, line width=2pt] (2.3,2.1) circle (15pt);
			\draw[red, line width=2pt] (2.3,-1.4) circle (15pt);
			
			\fill (3.0,1.8) circle (0.07);
			\fill (3.3,1.8) circle (0.07);
			\fill (3.6,1.8) circle (0.07);
			
			\fill (3.0,-1.8) circle (0.07);
			\fill (3.3,-1.8) circle (0.07);
			\fill (3.6,-1.8) circle (0.07);
		\end{tikzpicture}
		\vspace{-0.3cm}
		\caption{Conventional calibration process with a reference image using polynomial regression and \textit{Butterfly}-image from database introduced in \cite{Images}. Note the significant color correction from $x_{2}$ to $y_{2}$.}
		\label{color correction principle}
	\end{center}
	\vspace{-0.8cm}
\end{figure}

To address this issue, several calibration methods have been developed, which can be divided into two groups: algorithmic solutions and approaches using calibration targets such as ColorChecker boards \cite{ColorChecker}. While the latter require additional hardware, algorithmic methods are based on different mathematical principles, but follow the same general concept shown in Fig. \ref{color correction principle}. The pixel values of a reference image $x_{r}$ are extracted and used in the calibration algorithm to map the colors of one or more input images $x_{0}$,...,$x_{n}$ to the reference. The outputs $y_{0}$,...,$y_{n}$ are adjusted images, in which the color appears to be closer to the reference compared with the original input image. The existing calibration methods rely on one given reference image, that is randomly chosen. However, since it is possible that artifacts and color shifts appear in the reference image as well, this paper investigates a novel method that does not rely on a single reference image, but defines a consensus image $x_{c}$ resulting from the location parameter and other statistics, such as mean or median, thus producing the most realistic results. These statistical calculations are a common method and their effectiveness should be preferred over complicated and complex solutions. 

This paper is structured as follows. Section \ref{Related work} reviews previous work in the field of color calibration. In Section \ref{Proposed Method}, we describe our new approach. The results are presented in Section \ref{Tests and Results}, with detailed comparison to previous work. The summary and conclusion are given in Section \ref{Conclusion}.

\section{Related work}
\label{Related work}
%\vspace{-0.2cm}
To address the issue of color calibration, different solutions are possible. Color calibration using algorithmic approaches focuses on maintaining inter-camera color consistency by employing mathematical algorithms, while color correction approaches using calibration targets aim to ensure consistency between captured images and the real-world colors by utilizing physical reference targets.
%\vspace{-0.4cm}
\subsection{Algorithmic Approaches}
\label{Algorithmic Approaches}
%\vspace{-0.2cm}
Reinhard et al. \cite{meanAndVarianceReinhard} were one of the first to introduce a method for color calibration that uses global statistics in the uncorrelated color space in order to linearly match the colors of the source and reference images. However, in case an image requires non--linear color mapping, this approach fails due to the lack of the dominant non--linear statistics. 

Tian et al. \cite{histogramMatching} therefore proposed an approach called Histogram Matching, which uses the image histogram to determine the transformation matrices by linear models. Another method was introduced by Fecker et al. \cite{Fecker}, which is based on luminance and chrominance histogram matching. Since it was proven that results improve even more when multiple statistics and non--linear models are being explored, Oliveira et al. \cite{gaussianMixtureModel} made use of Gaussian mixture models to approximate the color histograms.

Further simple and efficient approaches are regression algorithms. Linear regression models, as introduced in \cite{linearRegression}, presuppose linear relationships between the input and the reference image, while the polynomial regression solution in \cite{polynomialRegression} offers an alternative by providing more flexibility in modeling complex relationships between image intensities. Instead of fitting a straight line, it fits a polynomial curve to the data, accommodating more intricate patterns. By increasing the degree of the polynomial, even higher improvements of color mappings can be achieved.

The idea in \cite{SIFTandAffine} is to make use of the Scale--Invariant--Feature--Transform in combination with affine transformations (SIFTcal) to detect corresponding key points. The obtained alignment and matching ensures a precise color registration and the transformation matrices for calibration are retrieved.

The principle of the Cross-Correlation-Model-Function (CCMF) \cite{crossCorrelationPorikli} involves calculating the cross-correlation matrix between two histograms. A model function is obtained to correct the input images to the reference image. 

\begin{figure*}[h!] 
	\begin{center}
		\begin{tikzpicture}[auto]
			\centering
			[node distance=2cm, >=Stealth, font=\scriptsize]
			
			% Nodes
			\node (Xn) at(0,-0.5) {\textbf{\textit{X}}$_{n}$};
			 \node(kreis)[draw, circle, fill=black, inner sep=1pt, right of = Xn] {};
			\node (algo) [right of=Xn, draw, fill=white, xshift=2.3cm, text width=2.8cm, align=center, minimum height=1.8cm] {(Weighted) mean (Weighted) median};
			\node (cal) [right of=algo, draw, fill=white, xshift=4.5cm, text width=1.6cm, align=center, minimum height=1.8cm, minimum width=2cm] {Calibration process};
			\node (randComp) [right of=cal, draw, fill=white, xshift = 3cm, text width=2cm, minimum height=1cm, align=center] {Comparison};
			\node (PSNR) [right of=randComp, xshift=1.5cm, yshift=0.3cm] {PSNR};
			\node (iCID) [right of=randComp, xshift=1.5cm, yshift=-0.3cm] {iCID};
			\node(reference)[above of = randComp, yshift = 0.5cm, text width=1.5cm, align=center]{Uncalibrated image};
			
			\draw[->] (kreis) -- ++(0,1.5) -| (cal.north);
			\draw[->] (Xn) -- (algo);
			\draw[->] (algo) -- node[above]{$x_{c,a}$, $x_{c,wa}$} node[below]{$x_{c,m}$, $x_{c,wm}$} (cal);
			\draw[->] (cal.east) -- node[above]{\textbf{\textit{Y}}$_{n}$}(randComp);
			\draw[->] (randComp) --++ (1.09,0.2) |- (PSNR);  % den hier noch nach oben
			\draw[->] (randComp) --++ (1.09,-0.1) |- (iCID);  % den hier noch nach unten
			\draw[->] (reference) -- (randComp);
			
		\end{tikzpicture}
		\vspace{-0.2cm}
		\caption{The pipeline from the input images \textbf{\textit{X}}$_{n}$ to the output images \textbf{\textit{Y}}$_{n}$ for the proposed calibration process. \textbf{\textit{X}}$_{n}$ and \textbf{\textit{Y}}$_{n}$ are representing all modified images from one image class of \cite{Images}.}
		\label{fig:CalibrationProcess}
	\end{center}
	\vspace{-0.6cm}
\end{figure*}
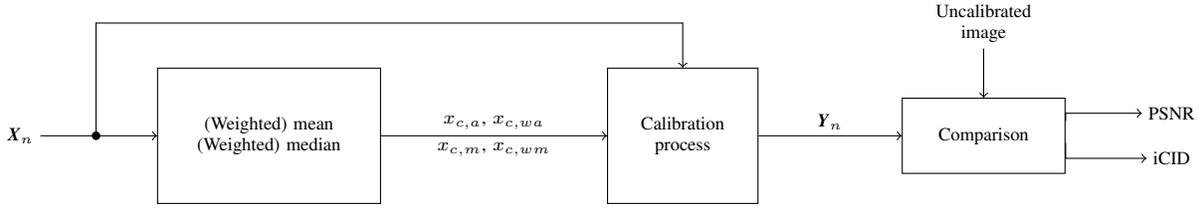

%\vspace{-0.4cm}
\subsection{Approaches using Calibration Targets}
\label{Alternative Approaches}
%\vspace{-0.2cm}
Since color calibration is important for achieving accurate and consistent results in imaging systems, not only algorithmic solutions but also approaches using calibration targets have been developed. Traditionally, reference targets, such as the ColorChecker or a gray wedge \cite{ColorChecker}, white reflectance boards \cite{ReflectanceStandards}, or the gray world assumption \cite{GrayWorldAssumption} have been employed as essential tools in calibration methodologies.

The ColorChecker \cite{ColorChecker} is a standardized chart with patches of known colors, used to evaluate and adjust the color reproduction capabilities of imaging devices. By capturing an image of the ColorChecker under the same lighting as the subject, differences between recorded and reference colors are calculated for precise calibration. Similarly, a gray wedge char with progressively darker or lighter gray patches, which is exemplary shown in Fig. \ref{fig:grayWedge}, provides a range of grayscale tones from 0 \% to 100 \% reflectance, essential for monochrome cameras. White reflectance boards \cite{ReflectanceStandards} serve as a white color reference under specific lighting, ensuring neutral white in captured images. However, natural images contain many distinct colors, which can reduce the effectiveness of these methods if the colors of interest are not represented in the calibration object. Additionally, these calibration targets focus on measurement consistency with reality, while our proposed method aims for inter-camera consistency.

The gray world assumption (GWA) \cite{GrayWorldAssumption} is based on color constancy, assuming the average scene color is neutral under neutral lighting. GWA corrects for lighting variations by adjusting image balance based on the perceived average color, achieving consistent and accurate color reproduction across different scenes and lighting conditions.

\vspace{-0.1cm}
\section{Proposed Method} \label{Proposed Method}
\vspace{-0.2cm}
When performing color calibration of multi-camera setups, one image is always taken as reference. However, we do not know which image displays the scene most realistically, thus a wrong reference with high deviation to the ground truth may be chosen. We propose a novel method to chose an optimum reference image for calibration using the principle of consensus images \cite{consensusImage} and refined location parameters \cite{locationParameter}, providing a robust foundation for improved color correction.

Our novel approach exploits different image statistics, specifically the combination of consensus images \cite{consensusImage} and refined location parameters \cite{locationParameter} and can be applied to almost all existing procedures. The idea has not been introduced before, but is based on simple principles, providing a robust and preferable foundation for improved color \-ca\-li\-bra\-tion. 

Consensus images offer a unique vantage point to uncover shared characteristics while diminishing the impact of aberrant values. Creating consensus images involves merging the individual images $x_{0},..,x_{N-1}$ pixel by pixel at their coordinates $(i,j)$ to create a new reference $x_{c}$ that captures shared patterns:
\begin{equation}
	x_{c}[i,j] = f(x_{0}[i,j],...,x_{N-1}[i,j]). \label{eq:Consensus}
\end{equation}
In this equation, $f(\cdot)$ denotes the function that is used to create the consensus image considering all available images $x_{n}$, $n=0,...,N-1$ of the camera array with $N$ cameras.

\begin{table*}[t!]
	\caption{Evaluation of the PSNR values of uncalibrated and calibrated database images \cite{Images} with synthetic distortion in comparison to the original image. A calibration to a random image $x_{r}$, a $x_{c,a}$, a $x_{c,wa}$, ia $x_{c,m}$ and a $x_{c,wm}$ was performed. The best results are marked in bold. A positive $\Delta$ indicates the improvement towards the uncalibrated image, a negative $\Delta$ indicates a deterioration.}
	\vspace{-0.2cm}
	\begin{center}
		\begin{tabular}{c|c|c:c|c:c|c:c|c:c|c:c}
			& \textbf{\parbox{0.9cm}{\centering uncal.}} & \textbf{\parbox{1.1cm}{\centering $x_{r}$}} & \textbf{\parbox{0.8cm}{\centering $\Delta_{r}$}} & \textbf{\parbox{1.1cm}{\centering $x_{c,a}$}} & \textbf{\parbox{0.8cm}{\centering $\Delta_{a}$}} & \textbf{\parbox{1.1cm}{\centering $x_{c,wa}$}} & \textbf{\parbox{0.8cm}{\centering $\Delta_{wa}$}} & \textbf{\parbox{1.1cm}{\centering $x_{c,m}$}} & \textbf{\parbox{0.8cm}{\centering $\Delta_{m}$}} &  \textbf{\parbox{1.1cm}{\centering $x_{c,wm}$}} & \textbf{\parbox{0.8cm}{\centering $\Delta_{wm}$}} \\
			\hline
			\hline
			\centering \textbf{Lin. Reg. \cite{linearRegression}} & \multirow{4}{*}{\centering 23.652} & 20.714 & -2.938 & 22.387 & -1.265 & 20.714 & -2.938 & \textbf{24.804} & 1.152 & 24.456 & 0.804 \\	
			\centering \textbf{Pol. Reg \cite{polynomialRegression}} &  & 19.513 & -4.139 & 22.098 & -1.554 & 19.506 & -4.146 & \textbf{23.882} & 0.2 & 23.419 & -0.233 \\
			\centering \textbf{SIFT \& Affine \cite{SIFTandAffine}} &  & 9.227 & -14.425 & 23.794 & 0.142 & 23.804 & 0.152 & 23.836 & 0.184 & \textbf{23.852} & 0.2\\	
			\centering \textbf{CCMF \cite{crossCorrelationPorikli}} & & 24.479 & 0.827 & 21.890 & -1.762 & 22.248 & -1.404 & \textbf{24.707} & 1.055 & 24.652 & 1.0 \\ \hdashline \noalign{\vskip 0.5ex}
			\centering \textbf{Average} & & 18.483 & -5.169 & 22.542 & -1.11 & 21.568 & -2.084 & \textbf{24.307} & 0.655 & 24.095 & 0.443 \\
		\end{tabular}
		\label{Table PSNR Results}
		\vspace{-0.5cm}
	\end{center}
\end{table*}

\begin{table*}[t!]
	\caption{Evaluation of the iCID values of uncalibrated and calibrated database images \cite{Images} with synthetic distortion in comparison to the original image. A calibration to a random image $x_{r}$, a $x_{c,a}$, a $x_{c,wa}$, ia $x_{c,m}$ and a $x_{c,wm}$ was performed. The best results are marked in bold. A negative $\Delta$ indicates an improvement towards the uncalibrated image, a positive $\Delta$ indicates a deterioration.}
	\vspace{-0.2cm}
	\begin{center}
		\begin{tabular}{c|c|c:c|c:c|c:c|c:c|c:c}
			& \textbf{\parbox{0.9cm}{\centering uncal.}} & \textbf{\parbox{1.1cm}{\centering $x_{r}$}} & \textbf{\parbox{0.8cm}{\centering $\Delta_{r}$}} & \textbf{\parbox{1.1cm}{\centering $x_{c,a}$}} & \textbf{\parbox{0.8cm}{\centering $\Delta_{a}$}} & \textbf{\parbox{1.1cm}{\centering $x_{c,wa}$}} & \textbf{\parbox{0.8cm}{\centering $\Delta_{wa}$}} & \textbf{\parbox{1.1cm}{\centering $x_{c,m}$}} & \textbf{\parbox{0.8cm}{\centering $\Delta_{m}$}} &  \textbf{\parbox{1.1cm}{\centering $x_{c,wm}$}} & \textbf{\parbox{0.8cm}{\centering $\Delta_{wm}$}} \\
			\hline
			\hline
			\centering \textbf{Lin. Reg. \cite{linearRegression}} & \multirow{4}{*}{\centering 14.893} & 21.022 & 6.129 & 16.774 & 1.881 & 21.010 & 6.117 & \textbf{12.082} & -2.811 & 12.506 & -2.387\\	
			\centering \textbf{Pol. Reg \cite{polynomialRegression}} &  & 22.561 & 7.668 & 17.476 & 2.583 & 22.700 & 7.807 & \textbf{13.232} & -1.661 & 13.667 & -1.226 \\
			\centering \textbf{SIFT \& Affine \cite{SIFTandAffine}} &  & 59.531 & 44.638 & 14.794 & -0.099 & 14.801 & -0.092 & 14.771 & -0.122 & \textbf{14.767} & -0.126 \\
			\centering \textbf{CCMF \cite{crossCorrelationPorikli}} & & 13.885 & -1.008 & 17.530 & 2.637 & 16.541 & 1.648 & 13.025 & -1.868 & \textbf{12.989} & -1.904 \\ \hdashline \noalign{\vskip 0.5ex}
			\centering \textbf{Average} & & 29.250 & 14.357 & 16.644 & 1.751 & 18.763 & 3.87 & \textbf{13.278} & -1.615 & 13.482 & -1.411 \\
		\end{tabular}
		\label{Table iCID Results}
		\vspace{-0.7cm}
	\end{center}
\end{table*}

An important part of this method lies in the utilization of location parameters, particularly the mean and median. They are common methods and are therefore well studied and the simplicity, effectiveness, and computational efficiency of these parameters are favored over complex alternatives. 

The arithmetic mean, calculated by summing up all values and dividing by the total count, measures a central tendency of a finite set of pixel values. The mean of $n$ images is derived using \eqref{eq: average}, where $i$ and $j$ correspond to the $i$--th row and $j$--th column of the images, pointing to one specific pixel.

\begin{equation}
	x_{c, a}[i,j] = \mu[i,j] = \frac{1}{N}\sum_{n=0}^{N-1}{x_{n}[i,j]} \label{eq: average}   
\end{equation}
In case of skewed distributions or high standard deviations $\sigma_{n}[i,j]$, with
\begin{equation}
	\sigma_{n}[i,j] = \sqrt{\frac{1}{N}\sum^{N-1}_{n=0} \bigl(x_{n}[i,j] - \mu[i,j]\bigr)^{2}}~~~~,
	\label{eq:standardDeviation}
\end{equation}
the mean is strongly affected by small and large values and might therefore not represent the consensus correctly, necessitating additional considerations, such as weights $w_{n}$ \eqref{eq:weightcalc}:
\begin{equation}
	w_{n}[i,j] = \frac{1}{1 + \sigma_{n}[i,j]} ~~~.
	\label{eq:weightcalc}
\end{equation}
The weighted mean consensus image $x_{c, wa}$ can then be generated according to % with the respective weights $w_{n}$.
\begin{equation}
	x_{c, wa}[i,j] = \frac{\sum_{n=0}^{N-1} w_{n}[i,j] x_{n}[i,j]}{\sum_{n=0}^{N-1} w_{n}[i,j]}~~~~,
	\label{eq:weightedAverage}
\end{equation}
where $x_{n}$ represents the pixel values of the $n$-th image,  while the weights $w_{n}(i,j)$ are calculated using \eqref{eq:weightcalc}.

A second possible approach for considering all available images for the calibration process is by forming their median. It is obtained by using a sorted list $\tilde{x}$. In case the number of entries is odd, the median is set to the middle value of $\tilde{x}$, for an even number of entries, the median is set to the arithmetic mean of both middle numbers, i.e. 
\begin{equation}
	x_{c, m}[i,j] = 
	\begin{cases}
		\tilde{x}_{\frac{N + 1}{2}}[i,j] & N\text{ odd}\\
		\frac{1}{2}\bigl(\tilde{x}_{\frac{N}{2}}[i,j] + \tilde{x}_{\frac{N+1}{2}}[i,j]\bigr) & N\text{ even} ~~~.
	\end{cases} 
	\label{eq: median}
\end{equation}
The median is robust against the influence of outliers, providing a reliable measure of central tendency. For further improvement, weights can also be added to the median calculation, resulting in a weighted median consensus image $x_{c, wm}$ \eqref{eq:weightedMedian}, which also takes the standard deviations of corresponding pixels across multiple images into account.% and thus minimizes the sum of weighted absolute differences. 
\begin{equation}
	\begin{split}
		x_{c,wm}[i,j] = \text{argmin}\biggl(&\sum^{N-1}_{n=0} w_{n}[i,j] \\
		&\bigl| x_{n}[i,j] - x_{c,wm}[i,j] \bigr|\biggr)
	\end{split}
	\label{eq:weightedMedian}
\end{equation}
$x_{c,wm}$ is iteratively determined to be the value that minimizes the absolute differences between all input images and the consensus image, making it the final weighted median value for the given set of images and corresponding weights.

%A third possible solution for forming a calibration image considering all available recordings is the mode, which can be calculated using
%\begin{equation}
%  x_{c,mode}(i,j) = {\text{argmax}}_{x_{n}} ~ \sum_{n = 0}^{N}I(x_{i},x_{j})
%  \label{eq:modus}
%\end{equation}
%\begin{equation}
%	x_{c,mode}(i,j) = {\text{argmax}}_{k} ~ \bigl(\sum^{N-1}_{n=0} I(x_{n}(i,j) = k)\bigr) ~~~~.
%	\label{eq:modus}
%%\end{equation}
%It is defined as the value occurring the most often within the samples. In the formula $I(\cdot)$ denotes an indicator function, which returns 1 if the expression within the parenthesis is true and 0 otherwise. In the context of image processing, this means that for each possible pixel value $k$, the sum of the indicator function across all images is calculated. Consequently, the value of $k$ is chosen that yields the maximum sum, which is the pixel value that is most frequently observed. In our tests we only consider 9 images, thus it is very unlikely, that two or more images have the same pixel value at the same position $(i,j)$. Nevertheless, other setups that consider more cameras than CAMSI might benefit from using $x_{c,mode}$ and could achieve promising results. 

The reference image $x_{r}$ is replaced by the consensus image $x_{c,a}$, $x_{c,m}$, or $x_{c,wa}$, $x_{c,wm}$ respectively, which is then used for the calibration algorithm. For calibration, different already existing algorithms $g(\cdot)$, such as linear or polynomial regression, can be used, mapping the color values of the input images $x_{0},...,x_{N-1}$ to the consensus image:
\begin{equation}
	x_{0}[i,j],...,x_{N-1}[i,j] \xrightarrow{\text{$g[i,j]$}} x_{c}[i,j] ~~~. \label{eq:mapping}
\end{equation}

%\vspace{-0.2cm}
\section{Experiments and Results} \label{Tests and Results} \vspace{-0.1cm}
After forming the consensus image, regression algorithms, color transformations and a cross-correlation model function algorithm are tested for camera array calibration. The methods themselves have not been modified, thereby preserving the original ideas of the authors.

For quantitative and qualitative evaluation of the efficiency of calibration methods, several well--established metrics exist. The Peak--Signal--to--Noise--Ratio (PSNR) \cite{PSNR} objectively measures the similarity between images by computing the ratio of the \-maxi\-mum possible pixel value to the mean square error (MSE, average of squared pixel differences between reference and modified images). A higher PSNR value indicates better color calibration accuracy. Since images often yield contrast, hue or saturation, which highly influence the PSNR value but barely affect the visual quality, an improved Color Image Difference (iCID) algorithm \cite{iCID} is available to evaluate the intuitive quality of calibration methods by measuring the perceptual pixel value difference between images. A smaller value indicates a more effective color calibration.
\begin{figure*}[h!]
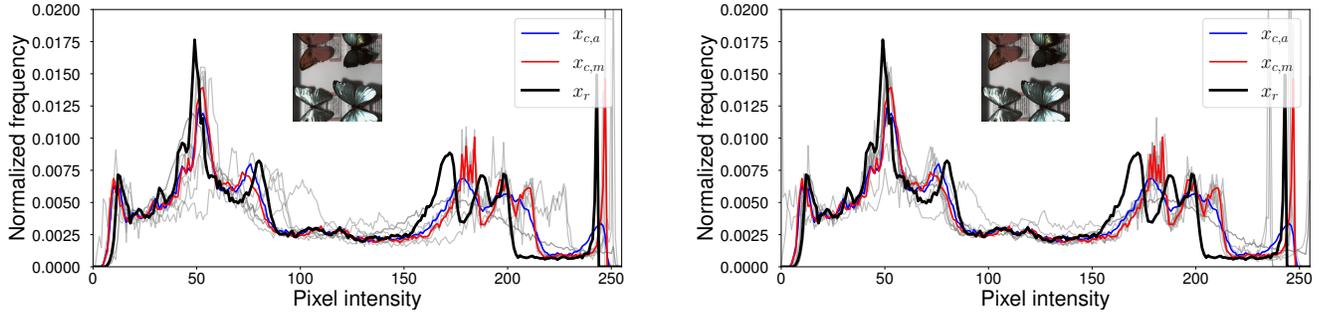
\vspace{-1.4cm}
	\begin{minipage}[b]{0.5\textwidth}
		\begin{tikzpicture}[      
			every node/.style={anchor=south west,inner sep=20pt},
			x=1mm, y=1mm,
			]   
			\node (fig1) at (0,0)
			{\resizebox{\linewidth}{!}{\input{Images_Paper/histogram_data_before_auswertung.pgf}}};
			\node (fig2) at (38,25)
			{\includegraphics[scale=0.09]{Images_Paper/Butterfly.png}};  
		\end{tikzpicture}
		\label{fig:subfigure3-a}
	\end{minipage}
	\hfill
	\begin{minipage}[b]{0.5\textwidth}
		\begin{tikzpicture}[      
			every node/.style={anchor=south west,inner sep=20pt},
			x=1mm, y=1mm,
			]   
			\node (fig1) at (0,0)
			{\resizebox{\linewidth}{!}{\input{Images_Paper/histogram_data_after_auswertung.pgf}}};
			\node (fig2) at (38,25)
			{\includegraphics[scale=0.09]{Images_Paper/Butterfly.png}};  
		\end{tikzpicture}
		\label{fig:subfigure3-b}
	\end{minipage}
	\vspace{-1.7cm}
	\caption{Histograms of database images \cite{Images} after calibration to random reference image $x_{r}$ (left) and to consensus image $x_{c,m}$ (right) using linear regression. The different reference images are marked in blue (mean consensus image), red (median consensus image) and black (random reference image) respectively.}
	\label{fig:hist dataset before and after calibration}
	\vspace{-0.2cm}
\end{figure*}
\begin{figure*}[h!]\vspace{-1.0cm}
	\begin{minipage}[b]{0.5\textwidth}
		\centering
		\begin{tikzpicture}[      
			every node/.style={anchor=south west,inner sep=20pt},
			x=1mm, y=1mm,
			]   
			\node (fig1) at (0,0)
			{\resizebox{\linewidth}{!}{\input{Images_Paper/histogram_CAMSI_before_auswertung.pgf}}};
			\node (fig2) at (40,35)
			{\includegraphics[scale=0.13]{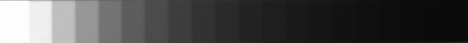}};  
		\end{tikzpicture}
		\label{fig:subfigure4-a}
	\end{minipage}
	\begin{minipage}[b]{0.5\textwidth}
		\centering
		\begin{tikzpicture}[      
			every node/.style={anchor=south west,inner sep=20pt},
			x=1mm, y=1mm,
			]   
			\node (fig1) at (0,0)
			{\resizebox{\linewidth}{!}{\input{Images_Paper/histogram_CAMSI_after_auswertung.pgf}}};
			\node (fig2) at (40,35)
			{\includegraphics[scale=0.13]{Images_Paper/Cam-1-klein.png}};  
		\end{tikzpicture}
		\label{fig:subfigure4-b}
	\end{minipage}
	\vspace{-1.7cm}
	\caption{Histograms of real CAMSI images after calibration to random reference image $x_{r}$ (left) and to consensus image $x_{c,m}$ (right) using linear regression. The different reference images are marked in blue (mean consensus image), red (median consensus image) and black (random reference image) respectively.}
	\label{fig:hist CAMSI before and after calibration}
	\vspace{-0.4cm}
\end{figure*}

In this paper, simulations with the color images of the \textit{TokyoTech} dataset introduced in \cite{Images} and real grayscale measurements with the CAMSI setup were performed. \textit{\-Tokyo\-Tech} contains publicly available multispectral images in the range between 420 nm and 650 nm (visible) and between 650 nm and 1000 nm (infrared). The image sizes are $512 \times 512$ pixels.

The \textit{TokyoTech} dataset was manually modified using random combinations of non--linear changes to reflect color inconsistencies across cameras. This included adding noise, enhancing single color channels, shifting color values, or changing the saturation and brightness of the images. Additionally the exposure and dynamic range were adjusted, what can be compared to different sensitivities of the sensors. Since the CAMSI-setup consists of 9 cameras, each of the database images was modified nine times, simulating recording errors and unequal color reproductions between the cameras. These adjustments were made randomly, which means that not every picture that represents a specific camera $n$ has undergone the same changes. During the modification, each RGB channel was treated individually. The nine adapted images of one of the original pictures were used to generate the consensus images $x_{c,a}$, $x_{c,m}$ and $x_{c,wa}$, $x_{c,wm}$ respectively. Those thus obtained images were then taken as references in the array calibration processes of the last four algorithms described in Section \ref{Related work}, while the modified images served as inputs requiring calibration. The calibrated results were afterwards compared to a randomly chosen evaluation image of the input set, expecting an improved outcome with respect to the comparison of the modified images. For better understanding, this process is visualized in Fig. \ref{fig:CalibrationProcess}.

The PSNR results of the conducted tests are shown in Table \ref{Table PSNR Results}, the iCID values are shown in Table \ref{Table iCID Results}. We highlighted the best result for each calibration algorithm in bold letters. First, we compared the modified images to a randomly chosen evaluation image to obtain a reference value for the following experiments. From the Tables \ref{Table PSNR Results} and \ref{Table iCID Results} we can infer that the comparison of the altered with the unaltered images results on average in a PSNR value of 23.65 dB and an iCID value of 14.89. %The results after a color calibration to a randomly chosen reference image are also listed. 
\begin{figure*}
	\begin{tikzpicture}
		\node(im1){\includegraphics[width=0.1\textwidth]{Images_Paper/Cam-2.png}};
		\node(im2)[right of=im1, xshift=1.0cm]{\includegraphics[width=0.1\textwidth]{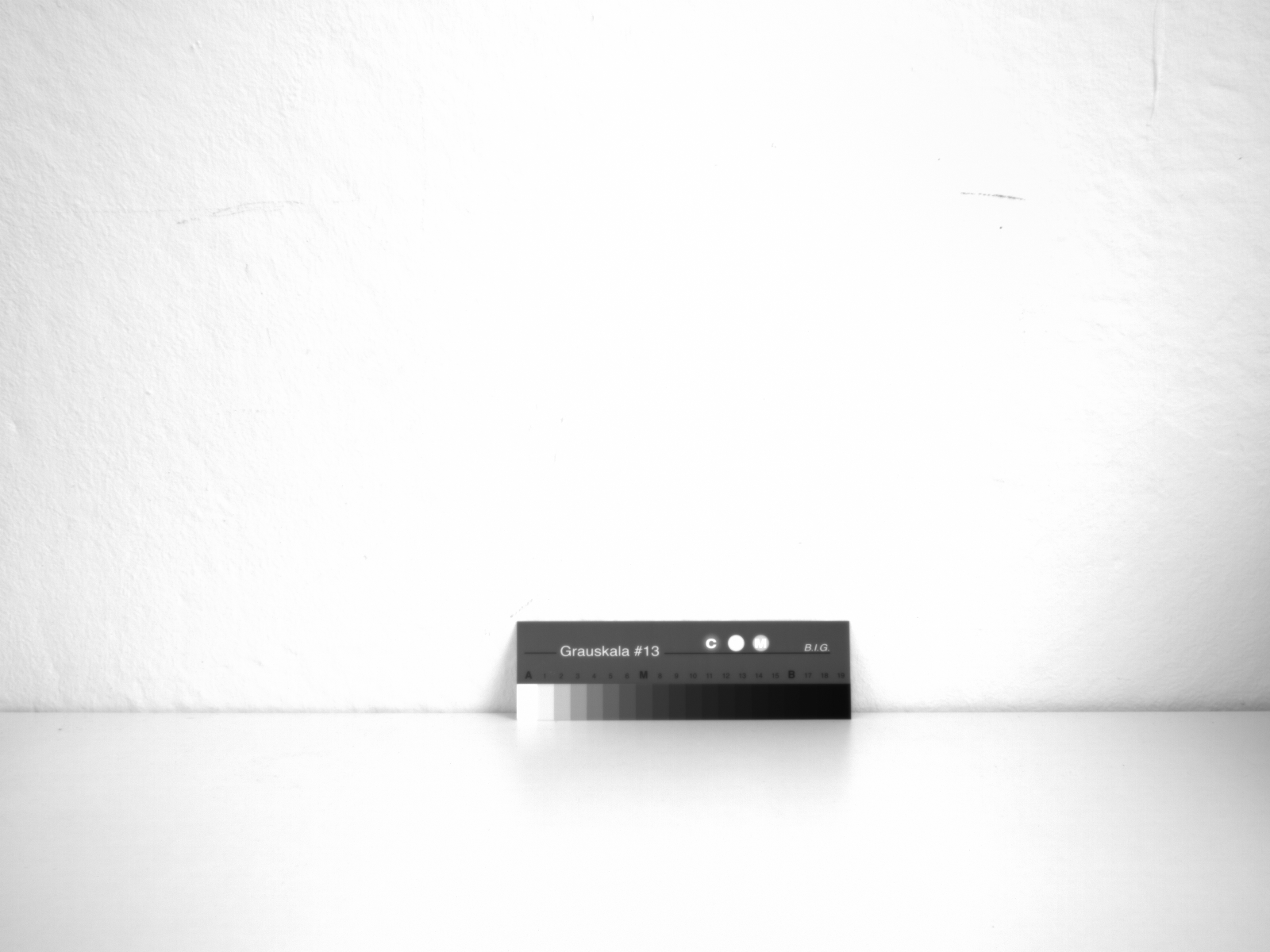}};
		\node(im3)[right of=im2, xshift=1.0cm] {\includegraphics[width=0.1\textwidth]{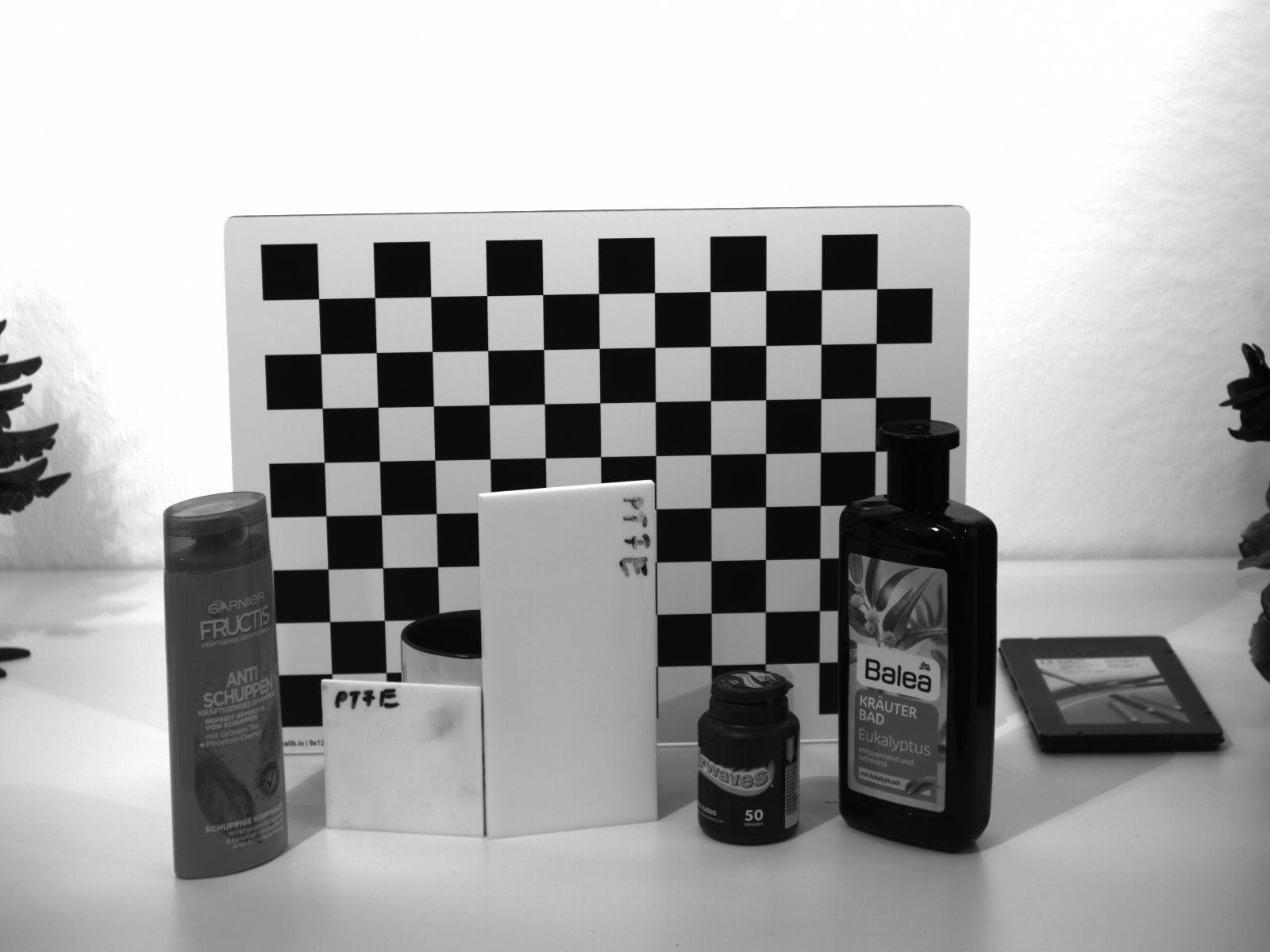}};
		\node(im4)[right of=im3, xshift=1.0cm]{\includegraphics[width=0.1\textwidth]{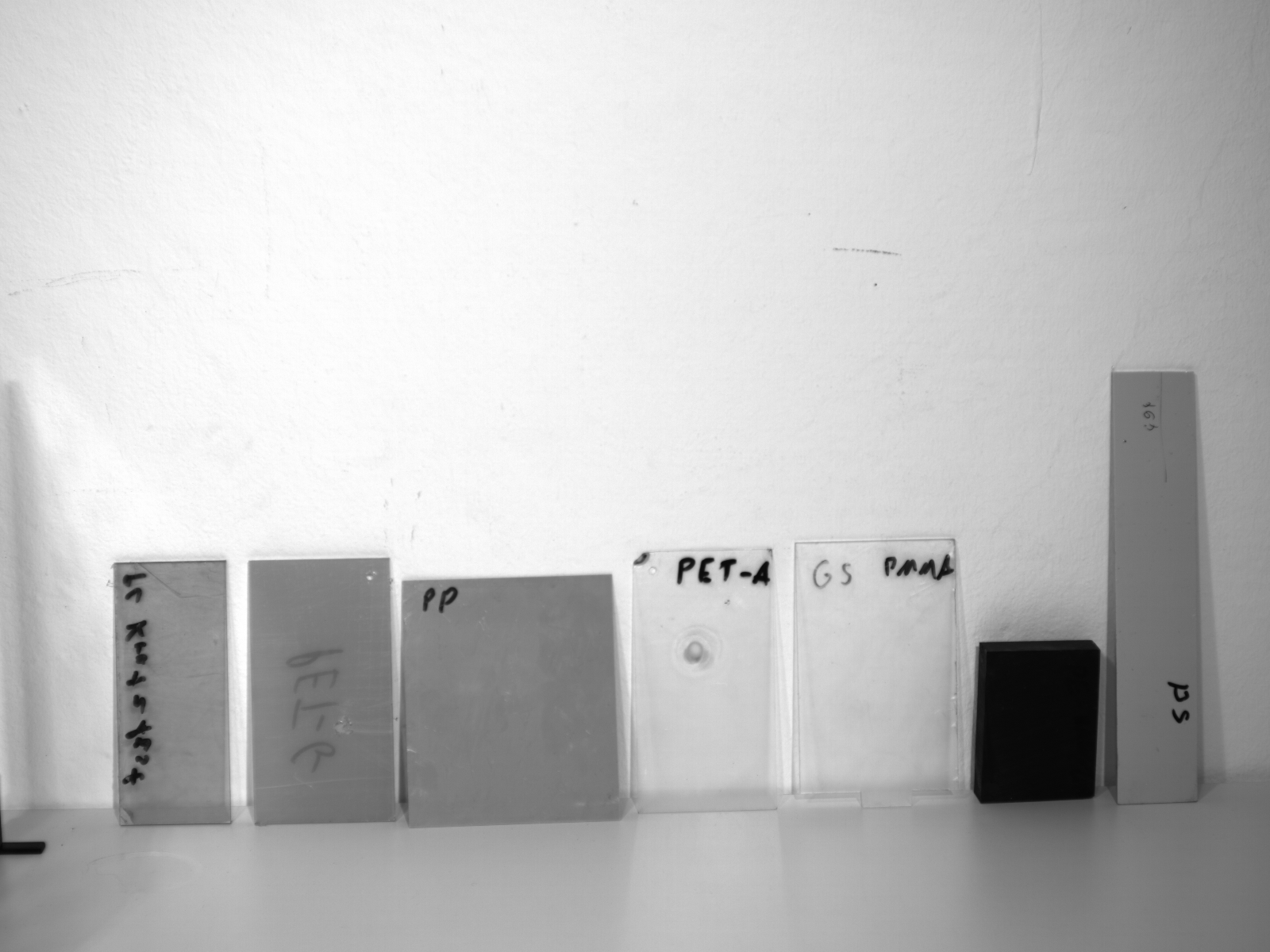}};
		\node(im5)[right of=im4, xshift=1.0cm]{\includegraphics[width=0.1\textwidth]{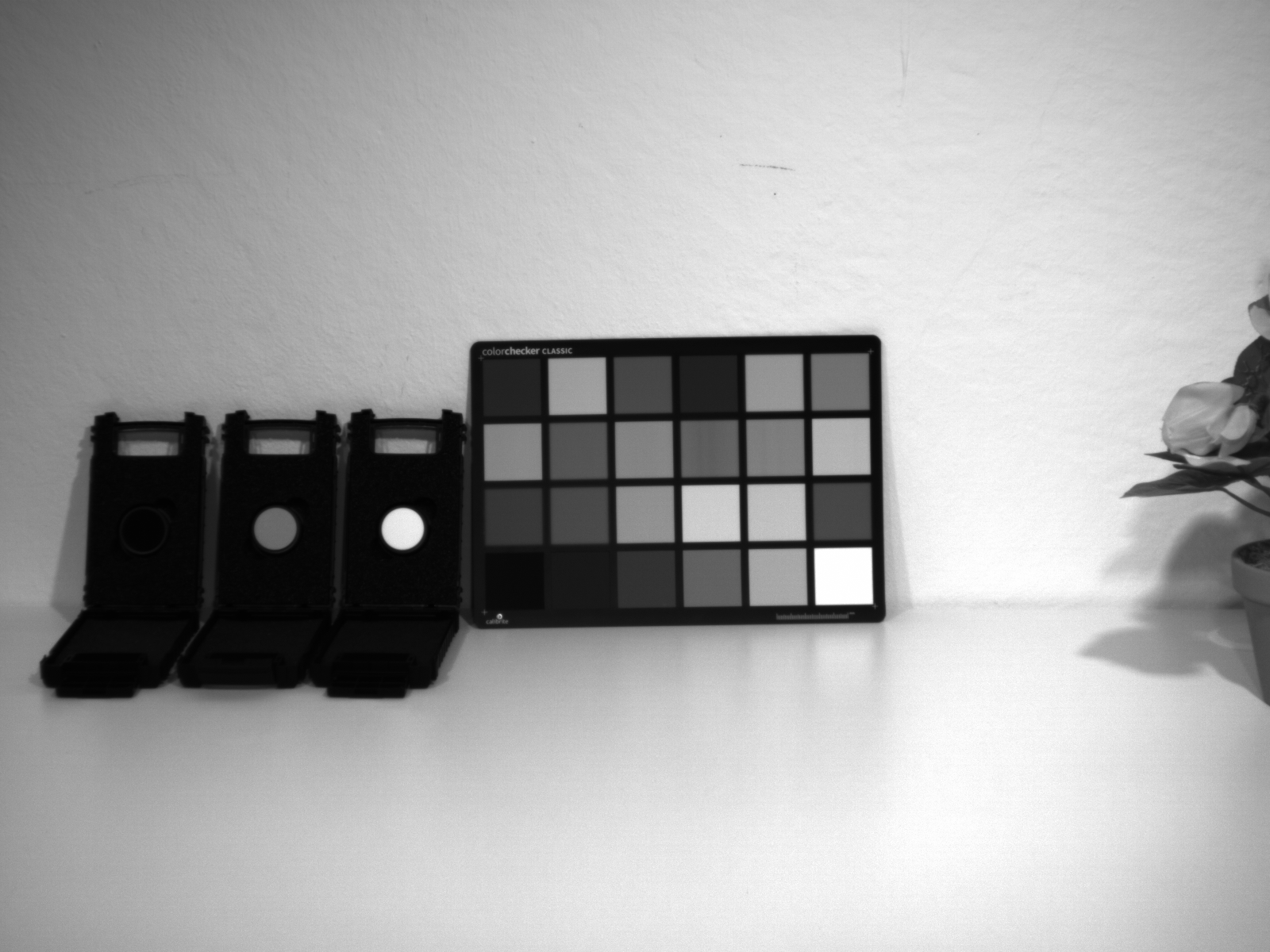}};
		\node(im6)[right of=im5, xshift=1.0cm]{\includegraphics[width=0.1\textwidth]{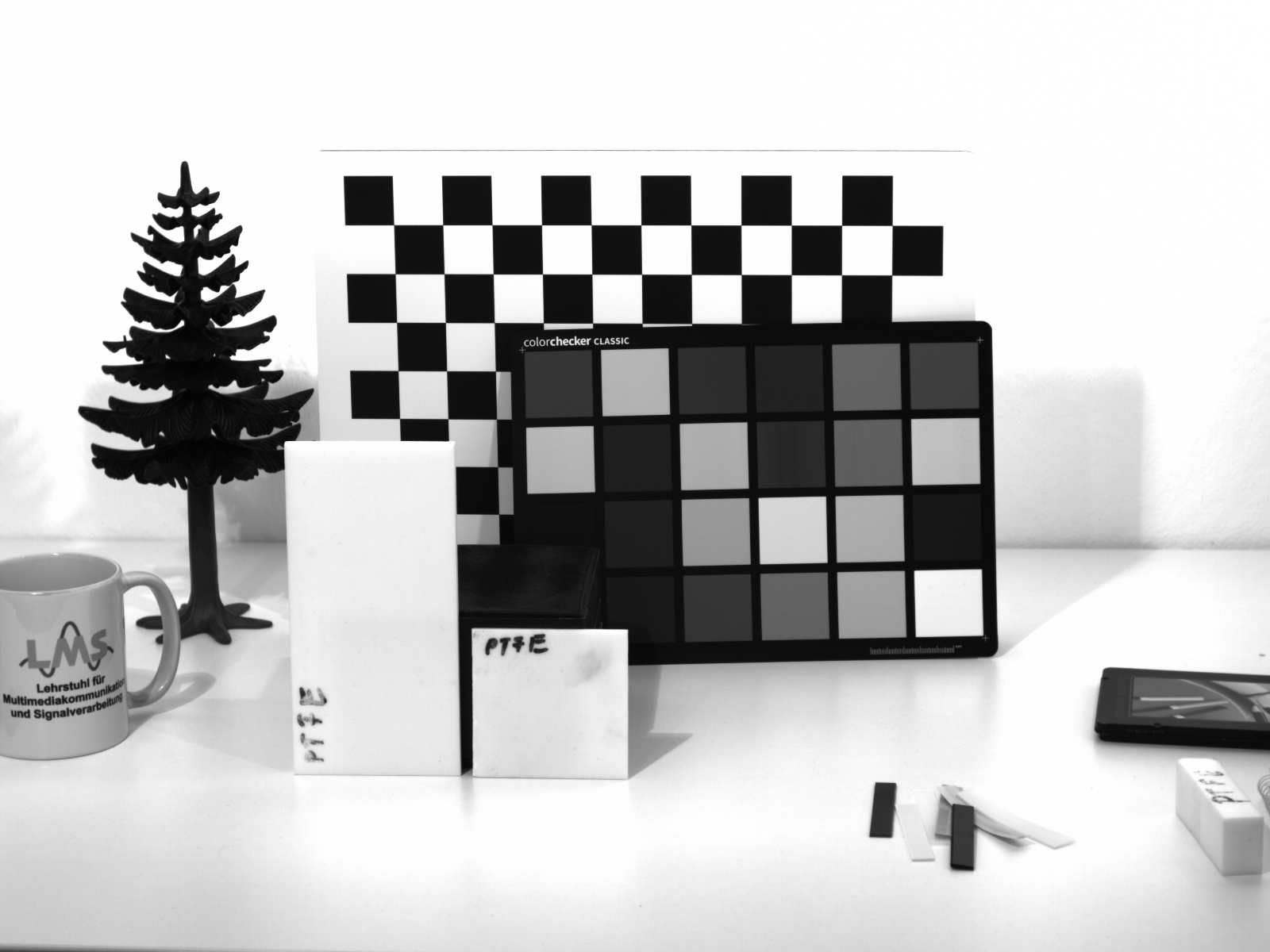}};
		\node(im7)[right of=im6, xshift=1.0cm] {\includegraphics[width=0.1\textwidth]{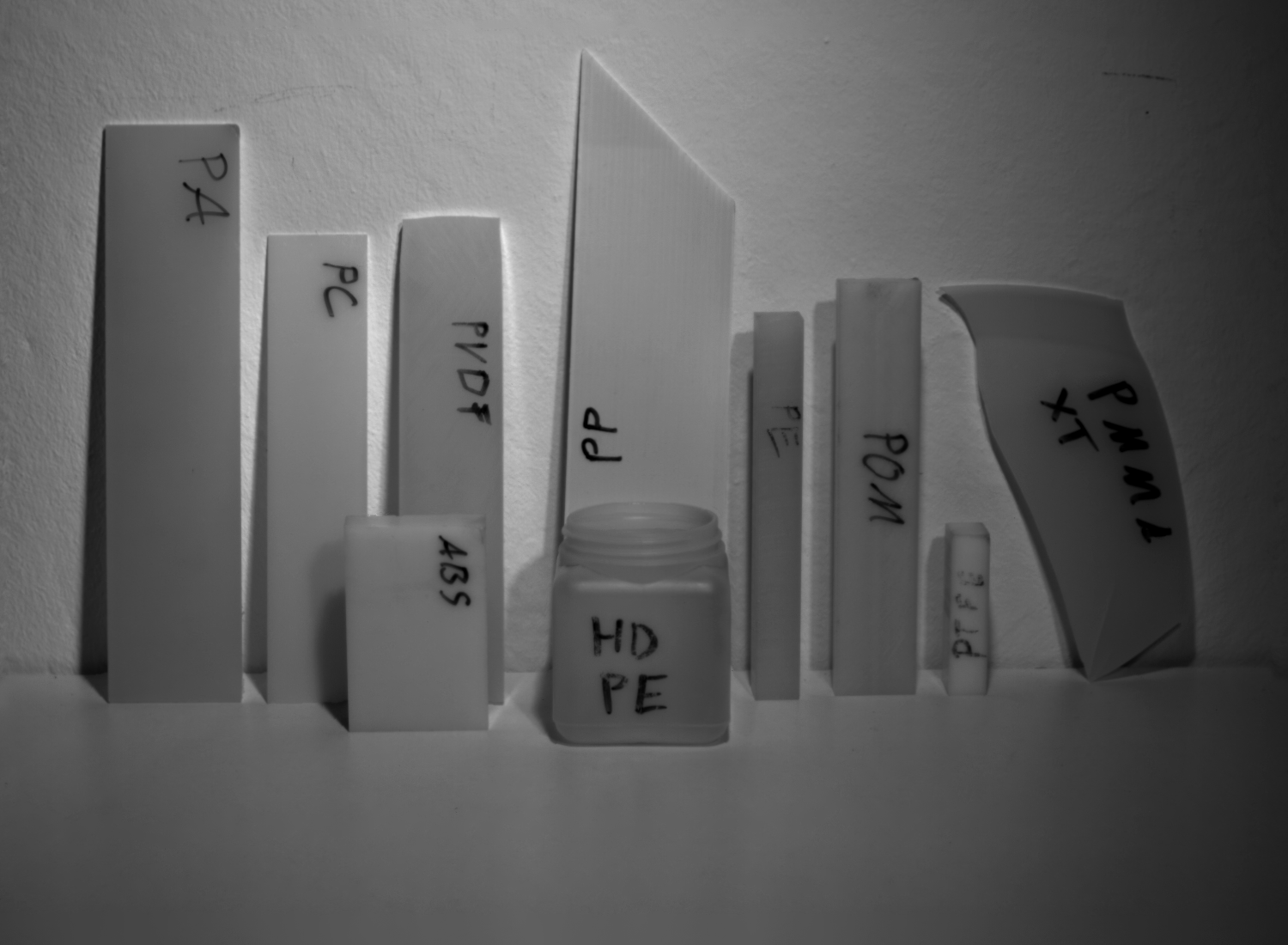}};
		\node(im8)[right of=im7, xshift=1.0cm]{\includegraphics[width=0.1\textwidth]{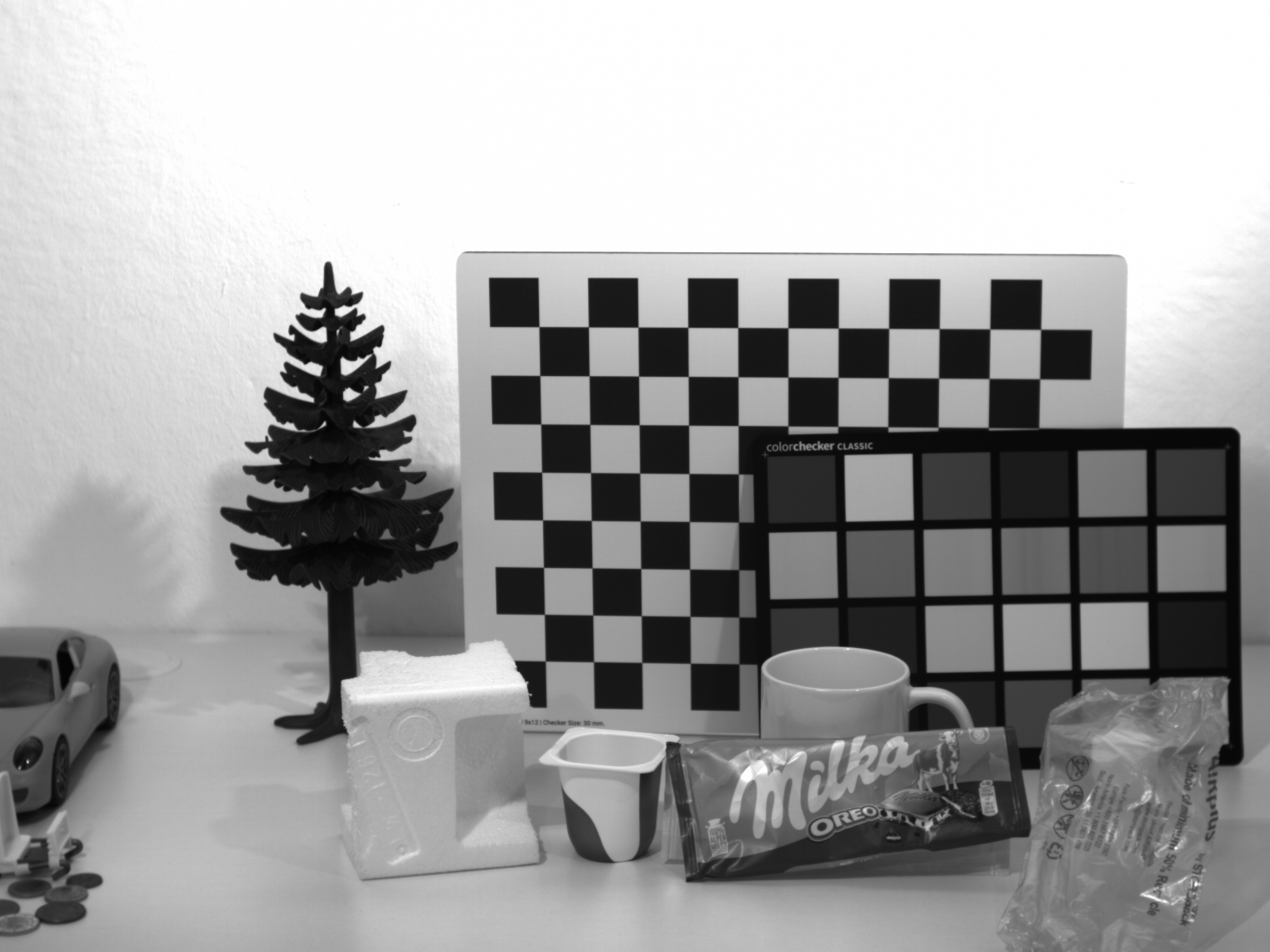}};
		\node(im9)[right of=im8, xshift=1.0cm]{\includegraphics[width=0.1\textwidth]{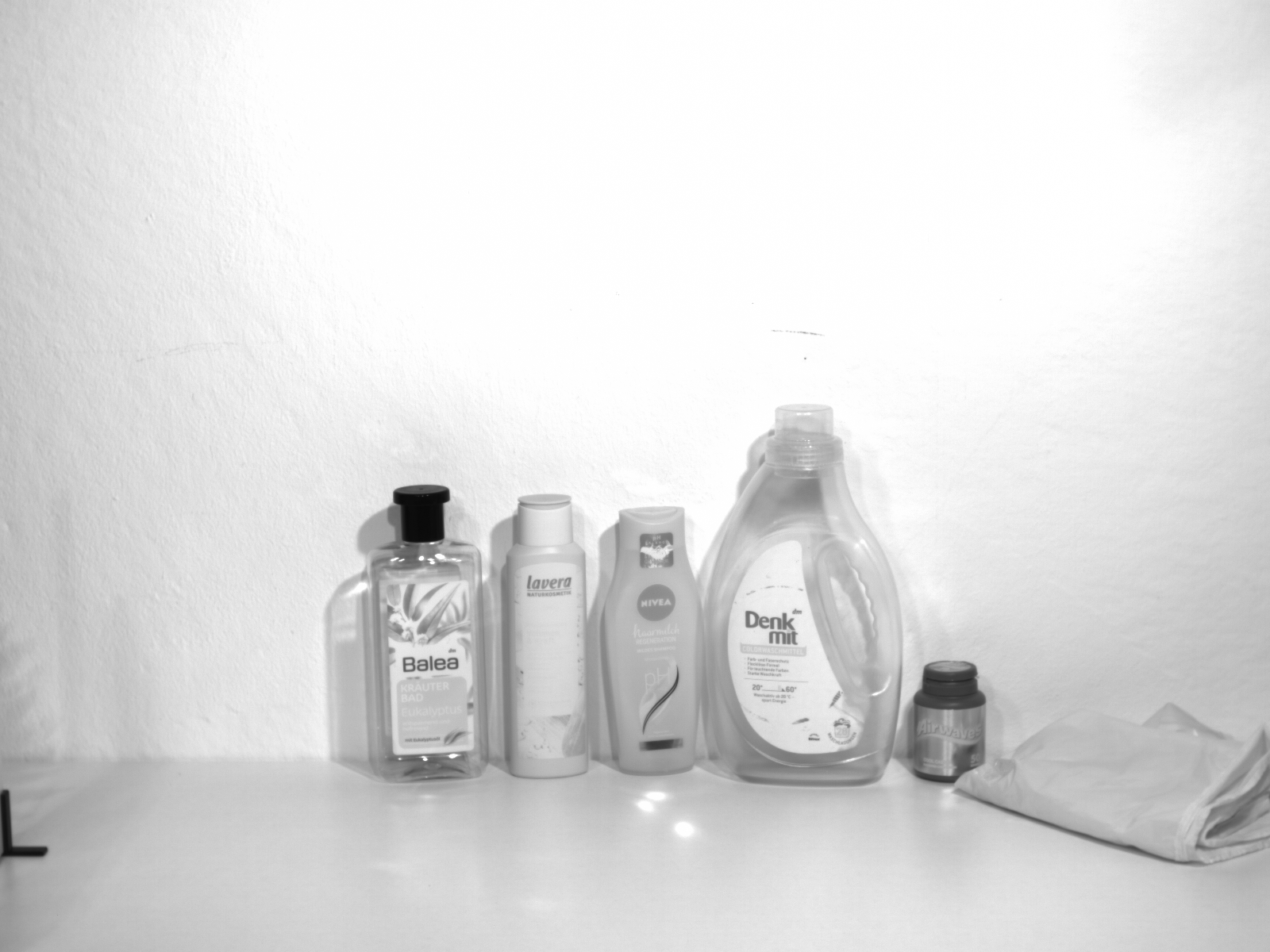}};
	\end{tikzpicture}
	\caption{Real scene images, recorded using CAMSI for further validation of our proposed method. Only one of the nine simultaneously recorded images is shown of each scene.}
	\label{fig:RealWorldImages}
	\vspace{-0.5cm}
\end{figure*}

When taking a closer look at the specific results of the linear regression calibration algorithm, we can see that color correction to a randomly chosen reference image leads to a decreased PSNR and increased iCID value. This is due to the fact that the randomly chosen calibration image contains high deviations to the other images through which a higher pixel difference arises. Our proposed approach addresses this precise problem by making use of all available information. However, when using the mean consensus image $x_{c,a}$ or $x_{c,wa}$ for color correction in the linear regression process, still a deterioration instead of improvement occurs. The results are of lower quality compared to a calibration to one of the median consensus images $x_{c,m}$ or $x_{c,wm}$. A possible explanation for this phenomenon is that, when forming the mean image, all images are equally weighted, thereby incorporating large deviations that strongly influence the result. Consequently, this leads to greater deviations between the images. Furthermore, from Tables \ref{Table PSNR Results} and \ref{Table iCID Results} we can infer that a color correction with the median consensus image $x_{c,m}$ results in the best quality for the linear regression calibration algorithm. This can be explained by the fact that single large deviations are not taken into account when forming the median. 

Throughout all tests, the results after a calibration to a median consensus image $x_{c,m}$ yields the highest PSNR and best iCID values. Nevertheless, for some calibration algorithms such as SIFTcal, we can improve these results even further by using the weighted median consensus image $x_{c,wm}$. This also proves that the simple and effective solution outperforms more complex and complicated algorithms.

Fig. \ref{fig:hist dataset before and after calibration} exemplary shows the histogram of of the \textit{Butterfly} color image of the tested database after calibration to a random reference image $x_{r}$ on the left hand side, and after calibration to the median consensus image $x_{c,m}$ using linear regression on the right hand side. The bold black line corresponds to the reference image $x_{r}$, the gray lines depict the individual camera recordings, demonstrating deviations between the images of the single cameras. The histogram of $x_{c,m}$ is displayed in red, while that of $x_{c,a}$ is shown in blue. As can be seen, the histograms of the calibrated images that were mapped to the consensus image $x_{c,m}$ are more aligned than after a calibration to a random reference $x_{r}$, supporting the results in Tables \ref{Table PSNR Results} and \ref{Table iCID Results}. This also proves that our proposed novel color calibration algorithm is robust against impacts such as color enhancement of single channels, noise, shifting color values, and saturation or brightness changes, as well as adjusted exposure and dynamic range.
	
For further validation, real images shown in Fig. \ref{fig:RealWorldImages}, were recorded and corrected using the CAMSI setup together with our proposed algorithm, which as well led to improved results. The histogram of the recorded gray wedge after calibration to one random reference image of the 9 available grayscale CAMSI camera recordings is shown in Fig. \ref{fig:hist CAMSI before and after calibration} on the left hand side, while on the right hand side the result after calibration to the median image $x_{c,m}$ using linear regression is displayed.

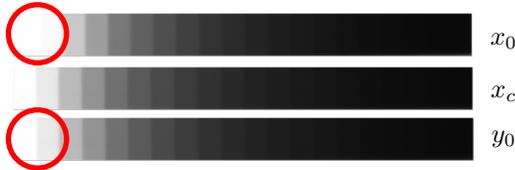
\begin{figure}
	\centering
	\begin{tikzpicture}	
		every node/.style={anchor=west ,inner sep=15pt},
		x=1mm, y=1mm,
		]  
		\node (fig3) at (0,0)
		{\resizebox{0.7\linewidth}{!}{\includegraphics{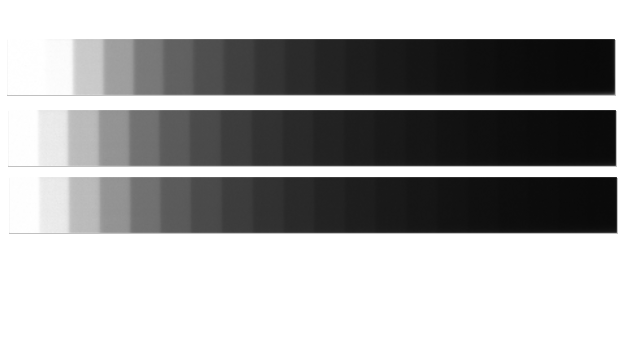}}};
		\node (x0) [right of=fig3,  xshift=2.5cm, yshift = 1.0cm]{$x_{0}$};
		\node (xc) [below of=x0,  yshift=0.3cm]{$x_{c}$};
		\node (y0) [below of=xc,  yshift=0.4cm]{$y_{0}$}; 
		
		\draw[red, line width=2pt] (-2.7,1.1) circle (11pt);
		\draw[red, line width=2pt] (-2.7,-0.3) circle (11pt);
	\end{tikzpicture}
	\vspace{-0.6cm}	
	\caption{Gray wedge recorded with CAMSI. The reflectances of the input image $x_{0}$ are hard to identify. After the calibration to the median consensus image $x_{c,m}$, the output image $y_{0}$ shows more distinguishable reflectances.}
	\label{fig:grayWedge}
	\vspace{-0.5cm}
\end{figure}

In Fig. \ref{fig:grayWedge} the input image $x_{0}$, the consensus image $x_{c}$ and the color corrected image $y_{0}$ of the gray wedge are shown, which were taken simultaneously under the same hardware and illumination conditions. Reconstruction and registration were performed to ensure almost perfect alignment of the individual images. Even in the visual example it can be seen that our proposed algorithm also performs well in a real world scenario, since the white colors of $x_{0}$ differ noticeably from $x_{c}$. After calibration, almost no deviation is visible and the different reflectance steps can be distinguished. The convergence of the CAMSI image histograms is evident, confirming the validity of our quality assessment, making this novel approach a usable method for color calibration for multispectral imaging when more than one camera at the same time is involved.

\vspace{-0.1cm}
\section{Conclusion}\label{Conclusion}
\vspace{-0.1cm}
In this paper we proposed a new method for color calibration of multi--camera setups. Instead of a random chosen reference image we make use of statistics to generate a consensus image of all cameras to calibrate the other images to. This way, we exclude the risk of using a bad reference image, since every camera output slightly differs from the perfect response. The tests have shown that the performance of four investigated color calibration algorithms (linear and polynomial regression, SIFTcal and cross-correlation model function) could be significantly improved and our novel solution works for both colored and monochromatic grayscale recordings. As future work, we will investigate if the proposed methods of color calibration for the visible spectrum also influence and improve the spectral camera response in the infrared range. Furthermore, we will explore machine learning calibration methods in connection with the proposed approach. 

\bibliographystyle{IEEEbib}
\bibliography{strings}

\pagebreak
\balance
\vfill

\end{document}